\documentstyle[11pt,epsfig]{article}
\renewcommand{\baselinestretch}{1.0}
\renewcommand{\appendix}
        {
        \par
        \setcounter{section}{0}
        \setcounter{subsection}{0}
        \gdef\afterthesectionpunctdefault{:}
        \gdef\thesection{{Appendix \Alph{section}}}
        \renewcommand{\theequation}{\Alph{section}\arabic{equation}}
        \setcounter{equation}{0}
        }

\begin{document}
\begin{center}
{\huge Uniform approximation for diffractive contributions}
{\huge  to the trace formula in billiard systems}

\vspace{1.0cm}

{\Large Martin Sieber$^{1,2}$, Nicolas Pavloff$^1$ and Charles 
Schmit$^1$}
\end{center}

\vspace{0.5 cm}

\noindent $^1$ Division de Physique Th\'eorique\footnotemark,
Institut de Physique Nucl\'eaire, F-91406 Orsay Cedex, France 
\hfill\break
\noindent $^2$ Institut f\"ur Theoretische Physik, Universit\"at Ulm,
D-89069 Ulm, Germany \hfill\break

\vspace{0.5 cm}
\begin{center}
{\bf Abstract}
\end{center}

	We derive contributions to the trace formula for the spectral
density accounting for the role of diffractive orbits in
two-dimensional billiard systems with corners.  This is achieved by
using the exact Sommerfeld solution for the Green function of a
wedge. We obtain a uniformly valid formula which interpolates between
formerly separate approaches (the geometrical theory of diffraction
and Gutzwiller's trace formula). It yields excellent numerical
agreement with exact quantum results, also in cases where other
methods fail.

\vspace{5cm}
\begin{math}
\footnotetext[1]{Unit\'e de Recherche des Universit\'es Paris
XI et Paris VI associ\'ee au CNRS}
\end{math}

\noindent PACS numbers:\hfill\break
\noindent 03.40.Kf ~ Waves and wave propagation: general mathematical
aspects.\hfill\break
\noindent 03.65.Sq ~ Semiclassical theories and 
applications.\hfill\break
\noindent 05.45.+b ~ Theory and models of chaotic systems.\hfill\break

\noindent IPNO/TH 96-22 ~ ULM-TP/96-3 \hspace{0.5cm}
{\it submitted to Physical Review E }\hfill\break
\newpage

\section{Introduction}

	Two-dimensional classical billiards became popular as model
systems exhibiting a rich variety of dynamical behaviour, ranging from
integrable to fully chaotic. Their quantum counterparts attracted much
interest starting in the 80's, both from the point of view of random
matrix theory and the semiclassical periodic orbit theory.  In the
latter approach one uses trace formulae of the type first derived by
Gutzwiller \cite{Gu90} and Balian and Bloch \cite{BB70,BB72}.

	During the last two years, following the route opened by Ref.
\cite{Vat94}, a number of studies (see Refs.\
\cite{Whe94,Pav95,Pri95,Bru95}) have concentrated on additional
contributions to the trace formula linked to diffractive effects near
regions where the classical Hamiltonian flow is discontinuous.  These
zones of discontinuity are known as ``optical boundaries" in the
literature. They lead to contributions from non-classical (so-called
diffractive) orbits hitting a corner of the billiard or creeping
around a smooth boundary.

 	Apart from the noticeable exception of Ref.\ \cite{Pri95} all
the work quoted above is based on Keller's ``geometrical theory of
diffraction" (GTD, see e.\,g.\ \cite{Keller}), i.\,e.\ on an extension
of geometrical optics which accounts for diffractive effects. Keller's
approach fails when the diffractive trajectory is very close to an
optical boundary, or equivalently when the diffractive orbit is close
to become an allowed classical trajectory (this will be clarified in
the text of the paper). In the present work we use a uniform
approximation for the Green function which does not have this
drawback. This allows to derive relatively simple formulae which are
uniformly valid. The method is applied to billiards whose boundary has
a slope discontinuity and thus we restrict our study to wedge
diffraction effects. To our knowledge there does not yet exist a
uniformly valid formula for the contributions of creeping orbits
(despite the progress made in Ref.\ \cite{Pri95}).

	The theory of uniform approximations for wedge diffraction has
 a long history which begins with a famous paper by Pauli
 \cite{Pauli}. In the late 60's and in the 70's the problem has been
 studied in detail. Much literature has been devoted to several types
 of approaches remedying the deficiency of the geometrical theory of
 diffraction. The approach most widely used is known as ``uniform
 asymptotic theory" and was developed by Ahluwalia, Boersma, Lewis and
 coworkers in Refs.\ \cite{Lew69,Ahl68,Ahl70,Lee76}. We have chosen
 here a technique more closely related to the original work of
 Sommerfeld and Pauli. It relies on an extension of the method of
 steepest descent due to Pauli which was carefully studied on a
 general setting by Clemmow \cite{Cle50}. The method, due to
 Kouyoumjian and Pathak, is known as ``uniform theory of diffraction"
 and is exposed in Refs.\ \cite{Kou74} and \cite{James}. Note that we
 apply the uniform approximation only to orbits with a single
 diffractive point. The treatment of multiple wedge diffraction is
 increasingly more involved as can be seen in work on double
 diffraction by half-planes (see Refs.\ \cite{Boe75,Lee75,Mitt78}) or
 wedges \cite{Sch91}. To our knowledge there does not exist to date a
 general uniform approximation for multiple wedge diffraction.

	The paper is organized as follows. In the next section we
recall the exact solution of the infinite wedge problem, derive a
uniform approximation for the Green function and compare it with the
result obtained from GTD. In Sec.\ 3 we use the Green function
obtained previously to derive contributions to the trace formula which
are uniformly valid. Readers mostly interested in the final result can
skip this part and go directly to Sec.\ 4, where we discuss the
previously obtained formula and several of its limits. In particular,
we show that this formula has the appealing feature of interpolating
between the semiclassical results of periodic orbit theory and the
formulae obtained in Refs.\ \cite{Vat94,Pav95,Bru95}. Sec.\ 5 contains
numerical applications for several simple billiard systems. In some
cases GTD gives reasonable results, but in other cases the uniform
approximation has to be used in order to describe the Fourier
transform of the spectral density correctly. Finally we discuss our
results and possible extensions in Sec.\ 6.

\section{The Green function of an infinite wedge}

	In this section we consider an infinite wedge of interior
angle $\gamma$ ($\gamma \in ]0,2\pi]$) with Dirichlet boundary
conditions and derive several approximations for the Green function.

	\subsection{The exact result}

	The exact solution of the problem was first given by
Sommerfeld for a wedge with $\gamma=2\pi$ (a half line) and an
incident plane wave, see \cite{Som}. The solution of the general
problem is easily inferred from his approach, a complete treatment is
given for instance by Carslaw in Refs.\ \cite{Car99,Car20}. We recall
here for completeness some properties of the solution.

	The Green function $G_\gamma(\vec{r},\vec{r}\,',E)$ of the
problem in dimensionless units is a solution of:

\begin{eqnarray}
\label{e0} (\Delta_{\vec{r}} + E) G_\gamma(\vec{r},\vec{r}\,',E) & = &
\delta (\vec{r} - \vec{r}\,' ) \; , \\ G_\gamma & \equiv & 0 \qquad
\hbox{if $\vec{r}$ or $\vec{r}\,'$ are on the boundary} \nonumber \; .
\end{eqnarray}

	Choosing a system of coordinates with the origin at the vertex
and the polar axis along one of the boundaries such that $\theta$ and
$\theta'$ are in $[0,\gamma]$ (see Fig.\ (2a)), one can write the
following integral representation for the exact solution:

\begin{equation}\label{e1}
G_\gamma(\vec{r} , \vec{r}\,',E) = g_\gamma (r,r',\theta'-\theta) -
g_\gamma (r,r',\theta'+\theta) \; , \end{equation}

\noindent with

\begin{equation}\label{e2}
g_\gamma (r,r',\phi_\sigma) = - {\displaystyle i\over\displaystyle
8\pi N} \int_{A+B} dz \; {\displaystyle H_0^{(1)} \left(
k\sqrt{\displaystyle r^2+r'^2-2 r r' \cos z}\, \right)
\over\displaystyle 1-\exp \left[ -i {\displaystyle
z-\phi_\sigma\over\displaystyle N} \right] } \; . \end{equation}

	In (\ref{e2}) and in the following the angles $\theta$ and
$\theta'$ always appear in the combination $\theta'\pm\theta$ and we
will denote $\phi_\sigma=\theta'-\sigma\,\theta$ ($\sigma = \pm
1$). Other quantities appearing in (\ref{e2}) are $N=\gamma/\pi$,
$k=\sqrt{E}$ which is the modulus of the wave-vector, $H_0^{(1)}$
which is the Hankel function of the first kind (see \cite{Abra}), and
$A$ and $B$ which are the contours in the complex plane drawn in
Fig.\,1.  In this figure one can further see the poles of the
integrand corresponding to $(z-\phi_\sigma)/N = 2 n \pi$ (with $n \in
\mbox{Z$\!\!$Z}$) -- they appear as black points -- and branch cuts
linked to the square root argument of $H_0^{(1)}$.  The shaded areas
are zones where the integrand increases without limit when one goes
away from the real axis (this is easily checked by using the leading
asymptotic term (\ref{e3c}) of the Hankel function). The integration
contour is quite arbitrary as long as it goes to infinity in the
indicated unshaded regions.
\begin{figure}[thb]
\begin{center}
\mbox{\epsfig{file=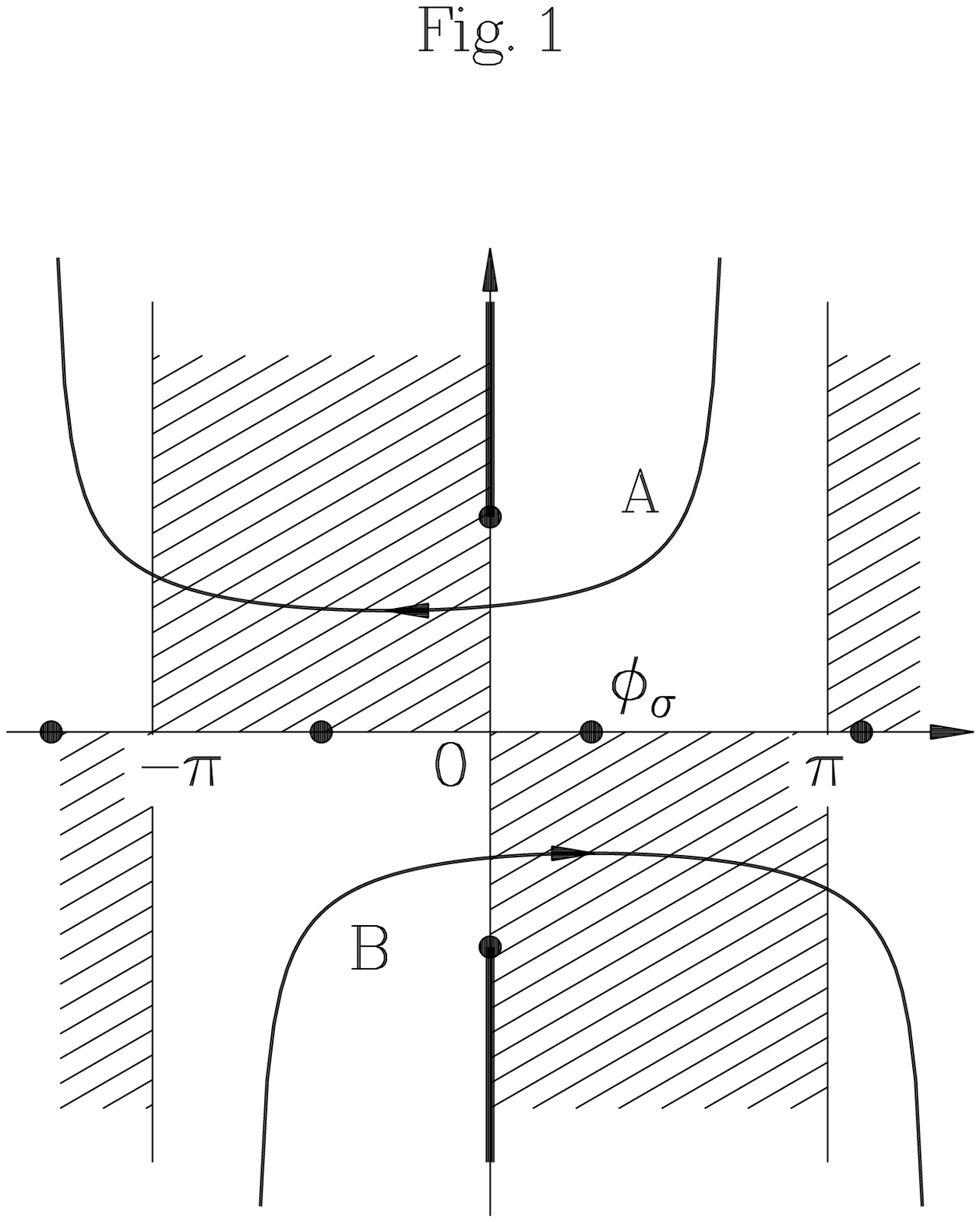,width=9cm,
bbllx=99pt, bblly=128pt, bburx=475pt, bbury=502pt,clip=}}
\end{center}
\caption{Integration contour in the complex plane for formula
(\protect\ref{e2}). The shaded areas are zones where the integrand
diverges when going away from the real axis. The black points are
poles and branch points of the integrand. The thick lines are
branch cuts.}
\end{figure}

	Essentially, $g_\gamma$ is a superposition of free Green
functions (with complex angles $z$). Considered as a function of
$\phi_\sigma$, it has perio\-dicity $2\pi N = 2\gamma$ and this
ensures that the Green function (\ref{e1}) satisfies the boundary
conditions for $\theta = 0$ and $\gamma$. By moving the contours $A$
and $B$ towards the real line $z \in [-\pi,\pi ]$ and taking into
account the poles of the integrand one obtains

\begin{equation}\label{e3a}
g_\gamma(r,r',\phi_\sigma) = -{\displaystyle i\over\displaystyle 4}
\left.\sum_n \right.^{\prime} H_0^{(1)} \left( k\sqrt{\displaystyle
r^2+r'^2-2 r r' \cos (\phi_\sigma-2n\gamma) }\, \right) +
h_\gamma(r,r',\phi_\sigma) \; , \end{equation}

\noindent where after a change of variable $h_\gamma$ can be written
in the form

\begin{equation}\label{e3b} h_\gamma(r,r',\phi_\sigma) =
{\displaystyle \sin (\pi/N)\over\displaystyle 8\pi N}
\int_{-i\infty}^{+i\infty} dz \; {\displaystyle H_0^{(1)} \left(
k\sqrt{\displaystyle r^2+r'^2 + 2r r' \cos z}\, \right) \over
\displaystyle \cos\left({\displaystyle z+\phi_\sigma\over\displaystyle
N}\right) - \cos (\pi/N) } \; .
\end{equation}

	The first term on the r.h.s.\ of (\ref{e3a}) contains the
contributions of those poles of the integrand of (\ref{e2}) which lie
between $-\pi$ and $\pi$; the prime indicates that the summation is
restricted to values of $n$ such that $-\pi \le \phi_\sigma-2n\gamma
\le \pi$. If $\phi_\sigma$ is exactly equal to $\pm\pi+2n\gamma$ then
the corresponding contribution to the summation has to be divided by
$2$. In (\ref{e3b}) the contour can be modified as long as no pole of
the integrand is crossed. A further requirement is that the part of
the contour extending to infinity has to start at $-i\infty$ with a
real part in $[0,\pi[$ and to extend to $i\infty$ with a real part in
$]-\pi , 0]$.

	The discrete summation in (\ref{e3a}) can be interpreted as
arising from allowed classical trajectories. For instance the case
$\phi_+=\theta'-\theta$ and $n=0$ gives a contribution $-(i/4)
H_0^{(1)} (k |\vec{r}-\vec{r}\,'|)$ and corresponds to the free
propagation from $\vec{r}\,'$ to $\vec{r}$. The other terms in the
summation correspond to trajectories experiencing specular reflections
on the boundaries (this is illustrated in Fig.\ (2b)).  If $\sigma=1$
(resp.\ $\sigma=-1$) the orbit has an even (resp.\ odd) number of
reflections.  These orbits correspond to successive applications of
the method of images and their contribution is known as the
geometrical term in the literature. When inserted back in Eq.\
(\ref{e1}) they give a term which will be denoted
$G_{geo}(\vec{r},\vec{r}\,',E)$ in the following.

	If the angle $\gamma$ is of the form $\pi/p$ ($p\in
\mbox{I$\!$N}^*$) then $\sin(\pi/N)=0$ and the term
$h_\gamma(r,r',\phi_\sigma)$ is zero: the geometrical term alone is
enough to fulfill the boundary conditions. This is due to the fact
that in this case the Green function can be determined by the method
of images. If $\gamma \neq \pi/p$, then $h_\gamma$ corresponds to the
contribution from diffraction. Hence the total Green function can be
written as a sum of a geometrical plus a diffractive term:

\begin{eqnarray}\label{e4a}
G_\gamma(\vec{r},\vec{r}\,',E) & = &
G_{geo}(\vec{r},\vec{r}\,',E)+G_{dif\!f} (\vec{r},\vec{r}\,',E) \; ,
\\ \hbox{with} \quad G_{dif\!f} (\vec{r},\vec{r}\,',E) & = &
h_\gamma(r,r',\theta'-\theta) - h_\gamma(r,r',\theta'+\theta) \;
.\nonumber \end{eqnarray}

\subsection{Geometrical theory of diffraction}

	We derive now a simple approximation for $G_{dif\!f} $. We
first replace the Hankel function by its asymptotic form for large
arguments (see \cite{Abra}):

\begin{equation}\label{e3c}
H_0^{(1)}(z) \approx \sqrt{{\displaystyle 2\over\displaystyle\pi z}}
\, \mbox{\Large e}^{\displaystyle i z-i\pi/4} \qquad \mbox{when}
\qquad |z|\gg 1 \; .\end{equation}

	The same approximation is used also in all the following for
the geometrical and diffractive Green functions, i.e.\ for all the
terms of Eqs.\,(\ref{e3a}) and (\ref{e3b}), the assumption being that
the distances measured along the paths (classical or diffractive)
going from $\vec{r}\,'$ to $\vec{r}$ are large compared to the
wavelength $\lambda=2\pi/k$. Then in the integral defining $h_\gamma$
there is a saddle point of the exponent at $z=0$ and a steepest
descent approximation yields:

\begin{equation}\label{e4} h_\gamma(r,r',\phi_\sigma) \approx
{\displaystyle 1\over\displaystyle 4\pi N} \, {\displaystyle
\sin(\pi/N) \over\displaystyle \cos(\phi_\sigma/N) - \cos(\pi/N) } \,
{\displaystyle \mbox{\Large e}^{ \displaystyle i
k(r+r')+i\pi/2}\over\displaystyle k\sqrt{\displaystyle r r'}} \;
.\end{equation}

	Incorporating this result into the expression (\ref{e4a}) for
$G_{dif\!f} $ one obtains a formula which can be cast into the form:

\begin{equation}\label{e7}
G_{dif\!f} (\vec{r},\vec{r}\,',E) \approx G_{sc}(\vec{r},\vec{r}_0,E)
{\cal D}(\theta,\theta') G_{sc}(\vec{r}_0,\vec{r}\,',E) \;
,\end{equation}

\noindent with

\begin{eqnarray}\label{e8}
{\cal D}(\theta,\theta') & = & {\displaystyle 2\over \displaystyle N}
\sin {\displaystyle \pi\over\displaystyle N} \left[ \left(
\cos{\displaystyle \pi\over N} - \cos{\displaystyle \theta
+\theta'\over\displaystyle N} \right)^{-1} - \left( \cos{\displaystyle
\pi\over N} - \cos{\displaystyle \theta -\theta'\over\displaystyle N}
\right)^{-1} \right] \\ \nonumber & & \\ \nonumber & = & -
{\displaystyle 4\over \displaystyle N} { \displaystyle \sin(\pi /N)
\sin (\theta /N) \sin (\theta' /N) \over \displaystyle
(\cos{\displaystyle \pi\over N} - \cos{\displaystyle \theta
+\theta'\over\displaystyle N} ) (\cos{\displaystyle \pi\over N} -
\cos{\displaystyle \theta -\theta'\over\displaystyle N} ) } \; .
\end{eqnarray}

	In (\ref{e7}) $G_{sc}$ is the free Green function evaluated
using (\ref{e3c}) and $\vec{r}_0$ is the point at the vertex (see
Fig.\ (2a)). Expressions (\ref{e7}) and (\ref{e8}) give the
diffractive part of the Green function in the ``geometrical theory of
diffraction" (see \cite{Keller}). They have the simple interpretation
as being the contribution of a (non-classical) diffractive trajectory
going from $\vec{r}\,'$ to $\vec{r}_0$ and then from $\vec{r}_0$ to
$\vec{r}$ (see Fig.\ (2b)). Using this approximation one can derive a
trace formula for the spectral density which accounts for diffractive
effects in the GTD approximation (see Refs.\
\cite{Vat94,Pav95,Bru95}).
\begin{figure}[thb]
\begin{center}
\mbox{\epsfig{file=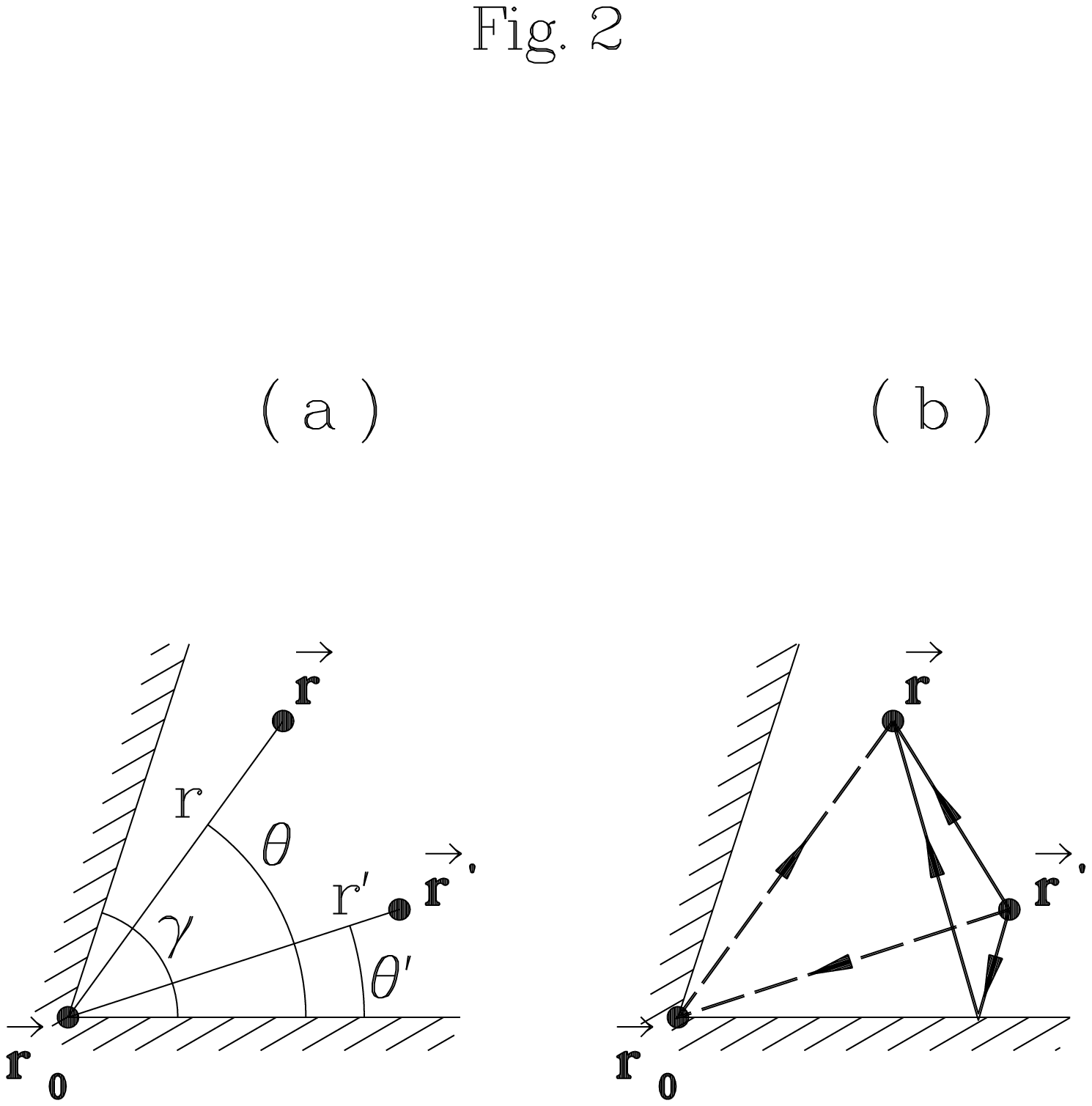,width=12cm,
bbllx=71pt, bblly=160pt, bburx=502pt, bbury=450pt,clip=}}
\end{center}
\caption{Figure (a) displays the notations used in the text.
Figure (b) shows two classical trajectories (solid lines) and
the diffractive orbit going from $\protect\vec{r}\,'$ to 
$\protect\vec{r}$ (dashed line).}
\end{figure}

	The quantity ${\cal D}(\theta,\theta')$ is known as the
diffraction coefficient. It is zero if $\theta$ (or $\theta'$) is
equal to 0 or $\gamma$, or if $\pi/\gamma$ is an integer. It diverges
on an optical boundary, i.\,e.\ if $\vec{r}$ and $\vec{r}\,'$ are such
that the diffractive orbit is the limit of a classical
trajectory. This is illustrated by the simple case of diffraction by a
sharp wedge ($\gamma >3\pi/2$) in Fig.\ 3. In this case there are two
optical boundaries represented by dashed lines. They correspond to
$\theta=\theta'+\pi$ (i.\,e.\ $\phi_+=\theta'-\theta=-\pi$) and
$\theta=\pi-\theta'$ (i.\,e.\ $\phi_-=\theta'+\theta=\pi$).  In the
terminology of geometrical optics, the first optical boundary
separates the illuminated and shadowed regions for direct rays when
the point $\vec{r}\,'$ is considered as a light source (boundary
between regions II and III), and the second optical boundary separates
the illuminated and shadowed regions for rays that are reflected on
one side of the wedge (boundary between regions I and II).  If
$\vec{r}$ lies near one of the optical boundaries then the diffractive
path is almost an allowed classical trajectory, and if $\vec{r}$ is
moved onto an optical boundary then the diffractive path coincides in
this limit with an allowed classical trajectory.
\begin{figure}[thb]
\begin{center}
\mbox{\epsfig{file=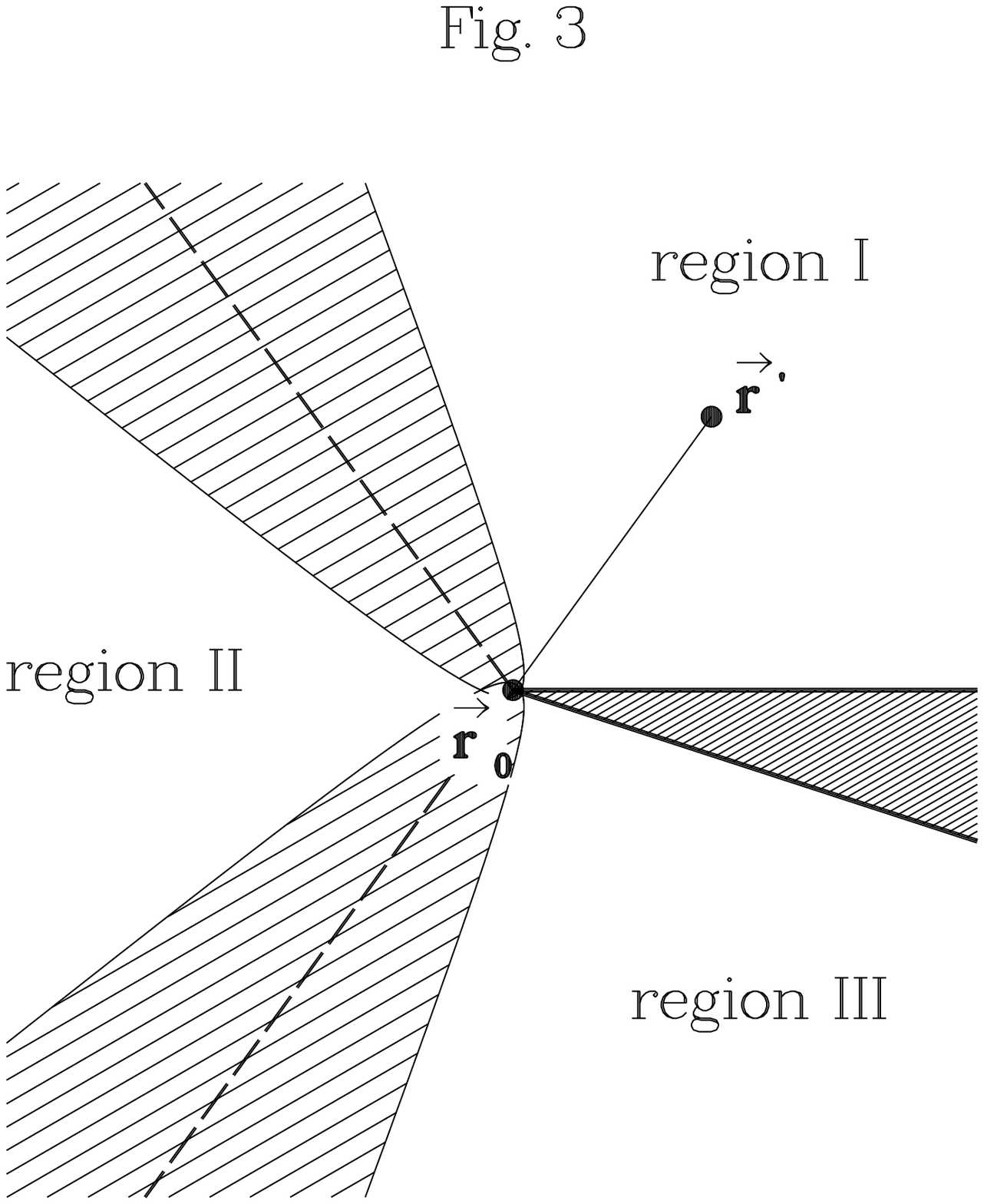,width=10cm,
bbllx=78pt, bblly=107pt, bburx=479pt, bbury=523pt,clip=}}
\end{center}
\caption{Optical boundaries (dashed lines) for an initial
point $\protect\vec{r}\,'$ in the case of diffraction by a
sharp wedge ($\gamma > 3\pi/2$). The transition regions
(for $r'=10 \lambda$) around the optical boundaries are
shaded.}
\end{figure}

	Looking in more detail at the origin of the divergence, one
sees from Eq.\ (\ref{e4}) that it occurs when there exits an integer
$n$ such that $\phi_\sigma = \pm\pi+2n N\pi$. In this case there is a
pole $z = \phi_\sigma \mp \pi-2n N\pi= 0$ in the integral
representation of the diffractive part (\ref{e3b}) which is at the
same position as the saddle point $z=0$ and thus the saddle point
approximation breaks down. More generally, the geometrical theory of
diffraction is only valid if all poles are sufficiently far away from
the saddle point $z=0$. This can be interpreted in terms of the
physical trajectories in the system because, in Sommerfeld's solution
(\ref{e1},\ref{e2}), the saddle point corresponds to the diffractive
orbit and the poles correspond to geometrical orbits.

\subsection{A uniform approximation}

	As seen above, one has to refine the steepest descent
evaluation of $h_\gamma$ in case that there is a pole of the integrand
near the saddle point $z=0$. This was first done by Pauli \cite{Pauli}
and we present here a slight modification of the original procedure
\cite{Kou74,James}. In a first step one can separate poles which are
possibly near one another by using the identity:

\begin{equation}\label{e8b}
{\displaystyle 2\sin(\pi/N)\over\displaystyle \displaystyle
\cos\left({\displaystyle z+\phi_\sigma\over\displaystyle N}\right) -
\cos (\pi/N) } = {\displaystyle
1\over\displaystyle\tan\left({\displaystyle z+\phi_\sigma
+\pi\over\displaystyle 2N}\right) } - {\displaystyle
1\over\displaystyle\tan\left({\displaystyle z+\phi_\sigma
-\pi\over\displaystyle 2N}\right) } \; .\end{equation}

	Hence $h_\gamma$ in Eq.\ (\ref{e3b}) can be rewritten as

\begin{equation}\label{e9} h_\gamma(r',r,\phi_\sigma) = 
u_{\gamma,+}(r',r,\phi_\sigma) - u_{\gamma,-}(r',r,\phi_\sigma) \;
                                  ,\end{equation}

\noindent where

\begin{equation}\label{e10} u_{\gamma,\eta} (r,r',\phi_\sigma) =
{\displaystyle 1\over\displaystyle 16\pi N} \int_{-i\infty}^{+i\infty}
 dz \; {\displaystyle H_0^{(1)} \left( k\sqrt{\displaystyle r^2+r'^2 +
 2r r' \cos z}\, \right) \over \displaystyle \tan\left({\displaystyle
 z+\phi_\sigma+\eta\,\pi\over\displaystyle 2N}\right) } \; ,
\end{equation}

\noindent and $\eta=\pm 1$ is a new index.

\

	If one denotes by $n_{\sigma,\eta}$ the nearest integer to
$(\phi_\sigma+\eta\,\pi)/(2\gamma)$ then $z=-(\phi_\sigma+\eta\,\pi) +
2\, n_{\sigma,\eta}\gamma$ is the pole of the integrand of (\ref{e10})
which is nearest to the saddle point $z=0$. Thanks to the separation
(\ref{e9}) the next pole in the integrand of (\ref{e10}) is at
distance $2\gamma$ and its effect can safely be neglected if $\gamma$
is not a small angle (this will be assumed in the
following). According to the method of Pauli one rewrites the
integrand by multiplying numerator and denominator by a function
imitating the behaviour of the original denominator but in which the
$z$ and $\phi_\sigma$ parts are separated. This procedure is not
unique, it corresponds to a specific choice of a uniform approximation
as will be discussed below. The choice for the function is
$\eta\,\sqrt{\displaystyle 2} \sin(z/2) + a_{\sigma,\eta}$, where
$a_{\sigma,\eta}$ is a measure of the separation between the saddle
point $z=0$ and the nearest optical boundary:

\begin{equation}\label{e12p}
a_{\sigma,\eta} = \sqrt{\displaystyle 2} \,
\cos({\displaystyle\phi_\sigma\over\displaystyle 2} -n_{\sigma,\eta}
\gamma) \quad\mbox{with}\quad
n_{\sigma,\eta}=\,\mbox{nint}\,[{\displaystyle\phi_\sigma+\eta\,\pi
\over\displaystyle 2\gamma}] \in \mbox{Z$\!\!$Z} \; .\end{equation}

	Using the asymptotic formula (\ref{e3c}) for the Hankel
function one obtains:

\begin{equation}\label{e11} u_{\gamma,\eta}(r,r',\phi_\sigma) \approx
{\displaystyle \mbox{\Large e}^{\displaystyle
-i\pi/4}\over\displaystyle 8\gamma\sqrt{\displaystyle 2\pi k}} \;
\int_{-i\infty}^{+i\infty}dz\; {\mbox{\Large e}^{\displaystyle i
k\sqrt{\displaystyle r^2+r'^2+2rr'\cos z}} \over
\displaystyle\eta\,\sqrt{\displaystyle 2} \sin(z/2) + a_{\sigma,\eta}
} \; F^{\sigma,\eta} (z) \; , \end{equation}

\noindent where $F^{\sigma,\eta} (z)$ is a smooth function at $z=0$,
even in the vicinity of an optical boundary (when $a_{\sigma,\eta}
\rightarrow 0$, see ({\ref{e15})):

$$F^{\sigma,\eta}(z)={\eta\,\sqrt{\displaystyle 2} \sin (z/2) +
a_{\sigma,\eta} \over\displaystyle(r^2+r'^2+2rr'\cos z)^{1/4} \,
\tan\left({\displaystyle z+\phi_\sigma+\eta\,\pi\over\displaystyle
2N}\right) } \; .$$

	Note that the integrands of (\ref{e11}) and (\ref{e10}) both
have the pole $z=-(\phi_\sigma+\eta\,\pi) + 2\, n_{\sigma,\eta}\gamma$
next to the origin as mentioned above.

	Now (\ref{e11}) is evaluated along the steepest descent path
at $z=0$ by a change of variable $z=\eta t \sqrt{2}\exp (i3\pi/4)$
with $t\in\mbox{I$\!$R}$ (the factor $\eta\sqrt{2}$ is here for
convenience). The smooth, non-singular part $F^{\sigma,\eta}$ of the
integrand is simply evaluated at $t=0$, and the phase of the
exponential function and the denominator are expanded in the vicinity
of the origin:

\begin{equation}\label{e12} \eta\,\sqrt{\displaystyle 2}\sin( z/2)  + 
a_{\sigma,\eta} = a_{\sigma,\eta}+t\,\mbox{\Large e}^{\displaystyle
3i\pi/4} + {\cal O}(t^3) \; , \end{equation}

\noindent and

\begin{equation}\label{e12z}
i k\sqrt{\displaystyle r^2+r'^2+2rr'\cos z} = i k ( r+r' ) -
{\displaystyle k\,r r'\over\displaystyle r+r'} \, t^2 + {\cal O}(t^3)
\; .\end{equation}

	Hence $u_{\gamma,\eta}$ is approximated by:

\begin{equation}\label{e13} u_{\gamma,\eta}(r,r',\phi_\sigma) \approx
{\displaystyle \mbox{\Large e}^{\displaystyle i k(r+r')- i
\pi/4}\over\displaystyle 8\gamma\sqrt{\displaystyle\pi k(r+r')}} \;
{\displaystyle a_{\sigma,\eta}\over
\displaystyle\tan\left({\displaystyle
\phi_\sigma+\eta\,\pi\over\displaystyle 2N}\right)} \;
\int_{-\infty}^{\infty}d t\; {\displaystyle \exp \left[-k r r'
t^2/(r+r')\right] \over\displaystyle t-a_{\sigma,\eta}\mbox{\Large
e}^{\displaystyle i\pi/4}} \; .\end{equation}

	After the expansion (\ref{e12}) of the denominator, the pole
in (\ref{e13}) is only approximately equal to the nearest pole in
(\ref{e10}), but they coincide when the pole approaches the stationary
point $t=0$. As noted above the choice of the uniform approximation
which leads to Eq.\ (\ref{e13}) is not unique (as discussed by
Clemmow, who calls it a ``partial asymptotic expansion", see
\cite{Cle50}). For example, another choice of a uniform approximation
can be obtained by making a change of variable which transforms the
exponent in (\ref{e11}) such that it becomes an exact quadratic
function, and multiply denominator and integrand by a function which
is linear in the new variable. Then one obtains Eq.\ (\ref{e13}) with
a different definition of $a_{\sigma,\eta}$ which is expressed in
terms of a ``detour parameter" such as used in the ``uniform
approximation theory'' (see Refs.  \cite{Lew69,Ahl68,Ahl70,Lee76}). In
practical applications, however, the differences between different
uniform approximations are small. We would like to add that the
uniform approximations can be further improved by including also
sub-leading terms of the asymptotic expansion of the Hankel function
and also higher order terms of the expansion of the integrand (see
Refs.\ \cite{Pauli,Cle50}). However, numerical checks show (see Fig.\
4 and below) that such refinements are not necessary here.

	We continue now with the integral (\ref{e13}) which can be
recognized as an integral representation of the modified Fresnel
function $K$ (see Appendix A and Eq.\ (\ref{f3})) and the final
expression for the uniform approximation for $u_{\gamma,\eta}$ is:

\begin{equation}\label{e14} u_{\gamma,\eta}(r,r',\phi_\sigma) \approx 
{\displaystyle 1\over\displaystyle 4N} {\displaystyle \mbox{\Large
e}^{\displaystyle i k(r+r')+i\pi/4}\over\displaystyle
\sqrt{\displaystyle \pi k(r+r')}} \; {\displaystyle |a_{\sigma,\eta}|
\over\displaystyle\tan\left({\displaystyle
\phi_\sigma+\eta\,\pi\over\displaystyle 2N}\right)} \; K \left(
|a_{\sigma,\eta} | \sqrt{ {\displaystyle k r r'\over\displaystyle
r+r'}} \, \right) \; .\end{equation}

	This expression remains finite on the optical boundary
$\phi_\sigma=-\eta\,\pi + 2\,n_{\sigma,\eta}\gamma$. As an optical
boundary is crossed, $a_{\sigma,\eta}$ goes through zero and changes
sign and one has:

\begin{equation}\label{e15} {\displaystyle |a_{\sigma,\eta}|
 \over\displaystyle\tan\left({\displaystyle
\phi_\sigma+\eta\,\pi\over\displaystyle 2N}\right)} \approx \eta
\sqrt{\displaystyle 2} N \; \mbox{sign} \,(a_{\sigma,\eta}) \qquad
\mbox{when} \qquad a_{\sigma,\eta}\rightarrow 0 \; .\end{equation}

	Hence although the problem of divergence has been eliminated
one arrives at a final form which is discontinuous. This was expected:
the exact terms (\ref{e3b}) and (\ref{e10}) already have this
behaviour; because of the separation (\ref{e4a}) of the total Green
function into a geometrical and a diffractive term, each contribution
($G_{geo}$ and $G_{dif\!f} $) is discontinuous at the optical
boundary, but their sum is continuous.

	As a r\'esum\'e of the results of this section we write down
the uniform approximation for the diffractive part of the Green
function which is a sum of four contributions:

\begin{equation}\label{e17}
G_{dif\!f} (\vec{r},\vec{r}\,',E) \approx {\displaystyle
1\over\displaystyle 4N} {\displaystyle \mbox{\Large e}^{\displaystyle
i k(r+r')+i\pi/4}\over\displaystyle \sqrt{\displaystyle \pi k(r+r')}}
\; \sum_{\sigma , \eta=\pm 1} {\displaystyle \sigma\eta \;
|a_{\sigma,\eta}| \over\displaystyle \tan\left({\displaystyle
\phi_\sigma+\eta\,\pi\over\displaystyle 2N}\right)} \; K \left(
|a_{\sigma,\eta}| \sqrt{ {\displaystyle k r r'\over\displaystyle
r+r'}} \,\right) \; ,\end{equation}

\noindent where $\phi_\sigma=\theta'-\sigma\, \theta$ ($\theta$ and
$\theta'$ being chosen in $[0,\gamma]$) and $a_{\sigma,\eta}$ is
defined in (\ref{e12p}).

	In the remaining part of this section we present some
numerical results illustrating the accuracy of the uniform
approximation and a failure of the GTD approximation. If the next
optical boundary is sufficiently far away one can replace the modified
Fresnel function in (\ref{e17}) by the first term of its asymptotic
expansion (\ref{f2}) and this leads to the GTD result
(\ref{e7},\ref{e8}). Roughly speaking, this approximation is good when
the argument of the $K$-function is greater than 3 and it fails when
the argument is less than 1.5. This puts a limit on the use of the
geometrical theory of diffraction illustrated in Fig.\ 3: inside the
dashed areas around the optical boundaries one has to use the uniform
approximation (these zones are known as ``transition region" in the
literature). The figure has been drawn for the case $r'=10\lambda$
($\lambda=2\pi/k$), and the transition regions are larger if one goes
to smaller values of $r'/\lambda$. In the limit $r\gg r'\gg\lambda$,
the transition width around an optical boundary at distance $r$ from
the apex is proportional to $r\sqrt{\lambda/r'}$ (relying on the
weaker assumption that $r, r'\gg\lambda$ one can show that it is
proportional to $[\lambda (r^2+r r')/r']^{1/2}$).  Outside of the
transition region expression (\ref{e7},\ref{e8}) is valid and shows
that $G_{dif\!f} $ is a small correction to $G_{geo}$. But near the
optical boundary the two terms are of the same order and exactly on
the boundary the two discontinuous contributions to $G_{geo}$ and
$G_{dif\!f} $ have exactly the same amplitude.

	The comparison between the uniform approximation (\ref{e17}),
the geometrical theory of diffraction (\ref{e7},\ref{e8}) and the
exact result (\ref{e3b},\ref{e4a}) for $G_{dif\!f} $ is made
quantitative in Fig.\ 4. In this figure one considers a wedge of
interior angle $\gamma=110^o$. The source point $\vec{r}\,'$ is fixed
at $\theta'=60^o$ and $r'=5\lambda$. The observation point $\vec{r}$
is at fixed distance from the vertex ($r=r'$) and $\theta$ scans the
interval $[0,\gamma]$. The modulus of $G_{dif\!f} $ is then plotted as
a function of $\theta$. In the figure one cannot distinguish the
uniform approximation from the exact result. The geometrical theory of
diffraction diverges on the optical boundaries (represented as dashed
lines in the upper part of Fig.\ 4).  Furthermore it is in clear
disagreement with the exact result for all values of $\theta$. Hence
one can infer that a trace formula based on Eq.\ (\ref{e7}) will not
correctly describe the spectrum in cases such as presented in Figure
4.
\begin{figure}[thbp]
\begin{center}
\mbox{\epsfig{file=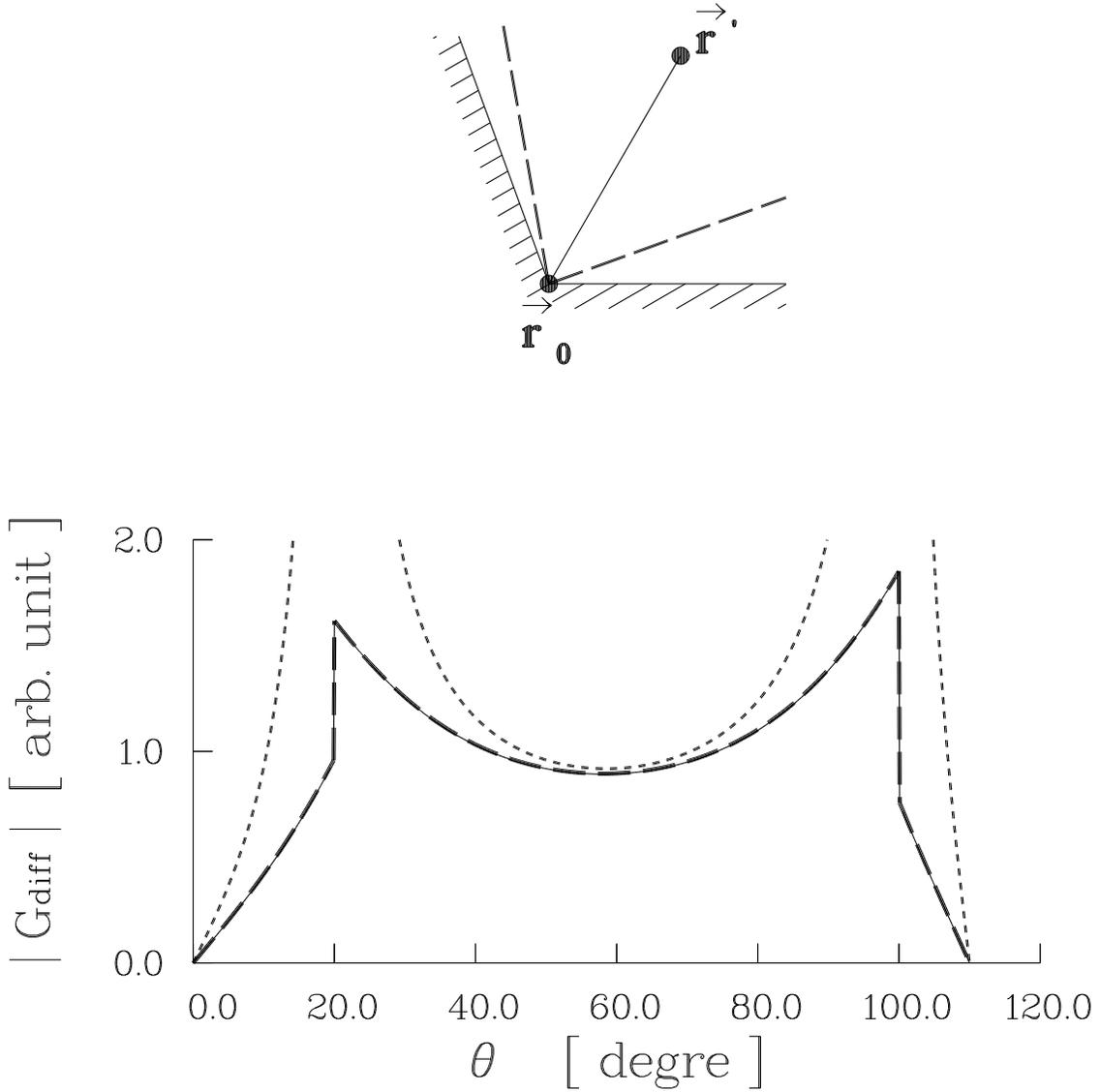,width=15cm,
bbllx=3pt, bblly=15pt, bburx=533pt, bbury=550pt,clip=}}
\end{center}
\caption{Modulus of 
$G_{dif\!f} (\protect\vec{r},\protect\vec{r}\,',E)$ for
fixed $\protect\vec{r}\,'$ and $r$ in a wedge with 
$\gamma=110^o$ ($r=r'=5\,\lambda$ and $\theta'=60^0$). 
$\theta$ scans the interval $[0,\gamma ]$. The upper part
of the figure displays the geometry considered, the optical
boundaries appearing as dashed lines.  In the lower part,
the solid line is the exact result 
(\protect\ref{e3b},\protect\ref{e4a}), the long dashed line
is the uniform approximation (\protect\ref{e17}) and the
short dashed line is the GTD result 
(\protect\ref{e7},\protect\ref{e8}).}
\end{figure}

\section{Diffractive orbits in the trace formula}

We consider now a closed two-dimensional region ${\cal B}$ with a
boundary $\partial {\cal B}$ smooth everywhere except at a finite
number of points where its slope is discontinuous.  The spectral
density $d(k)$ of this system has semiclassical contributions from
periodic orbits as well as from diffractive orbits. The latter ones
are closed orbits which have a finite number of points on vertices of
the billiard (we call these points diffractive or corner points in the
following) and follow the law of geometrical optics between two
diffractive points.  Within the framework of the geometrical theory of
diffraction, the contribution of a diffractive orbit $\xi$ to the
level density has been derived in \cite{Vat94,Pav95,Bru95} and it is
given by

\begin{equation} \label{z0}
d_\xi(k) = \frac{L_0}{\pi} \left[ \prod_{i=1}^p \frac{{\cal D}_i }{
\sqrt{8 \pi k (M_i)_{12}}} \right] \cos(k L - \frac{\pi}{2} \nu -
\frac{3 \pi}{4} p ) \; .
\end{equation}

Here $L_0$ and $L$ are the primitive and total length of the
trajectory, respectively, $p$ is the number of diffractive points,
${\cal D}_i$ is the diffraction coefficient in the $i$-th corner
(cf. Eq.\ (\ref{e8})), $(M_i)_{12}$ is the (12)-element of the
stability matrix at unit energy for a part of the trajectory between
two corners, and $\nu$ is the number of conjugate points plus twice
the number of reflections on the boundary between corners.

	According to (\ref{z0}) each corner point decreases the
contribution of a diffractive orbit by an order ${\cal
O}(k^{-1/2})$. This is correct only if the diffractive trajectory is
sufficiently far away from the optical boundaries in every corner
point. In the opposite case that the trajectory lies on an optical
boundary in every corner point it can be shown that its contribution
is of the same order in $k$ as that of a regular periodic orbit (see
below). In the following we will go beyond the GTD approximation and
derive a contribution to $d(k)$ from diffractive orbits with one point
in a corner that interpolates between these two regimes. We will use
for this purpose a method of uniform approximation similar to that
exposed in the previous section.

The starting point of our derivation is the boundary element
method. It is a reformulation of the quantum mechanical eigenvalue
problem in terms of a Fredholm equation of the second kind for the
normal derivative of the wave function on the boundary (see e.\,g.\
Refs.  \cite{Ber84,Bog92,Har92,Alo94,Ste} for discussion and
application in the context of the trace formula). More specifically if
we denote by $\vec{r}(s)$ a point of the boundary with curvilinear
abscissa $s$ and by $u(s)$ the normal derivative of the wave function
at this point, one has the following integral equation for the case of
Dirichlet boundary conditions:

\begin{equation}\label{z1} u(s') = - 2 \int_{\partial {\cal B}} d s \,
 u(s) \; \partial_{\hat{n}'} G_0(\vec{r},\vec{r}\,',E) \;
,\end{equation}

\noindent where $\vec{r}=\vec{r}(s)$, $\vec{r}\,'=\vec{r}\,'(s')$,
$\hat{n}'$ is the outward normal vector to $\partial {\cal B}$ at
point $\vec{r}\,'$, $\partial_{\hat{n}'}$ is the projection of the
gradient onto $\hat{n}'$, and $G_0(\vec{r},\vec{r}\,',E)=-(i/4)
H_0^{(1)}(k |\vec{r}-\vec{r}\,'|)$ is the free Green function. The
integral relation (\ref{z1}) has non-vanishing solutions $u(s)$ only
if

\begin{equation}\label{z2} \mbox{det}\; (\hat{I} - \hat{Q}(k)) = 0 \; 
,\end{equation}

\noindent where $\hat{I}$ is the identity and $\hat{Q}(k)$ is an
integral operator which, when applied to the function $u(s)$, gives
the r.h.s.\ of Eq.  (\ref{z1}). The zeros of (\ref{z2}) are the exact
quantum energies of the system and the oscillatory part $\tilde{d}(k)$
of the level density can be expressed as

\begin{equation}\label{z4} \tilde{d}(k) =
-{\displaystyle 1\over\displaystyle\pi} \, \Im {\displaystyle
d\over\displaystyle d k} \ln \, \mbox{det}\, (\hat{I} - \hat{Q}(k)) =
{\displaystyle 1\over\displaystyle\pi} \, \Im \, \sum_{n=1}^{\infty}
{\displaystyle 1 \over\displaystyle n} \; {\displaystyle
d\over\displaystyle d k} \left[ \, \mbox{Tr}\, \hat{Q}^n(k) \right] \;
,\end{equation}

\noindent with

\begin{equation}\label{z5} \mbox{Tr}\, \hat{Q}^n(k) = (-2)^n
\int_{\partial {\cal B}} ds_1 ... ds_n \, \partial_{\hat{n}_1}
G_0(\vec{r}_2,\vec{r}_1,E) \, \partial_{\hat{n}_2}
G_0(\vec{r}_3,\vec{r}_2,E) \dots \partial_{\hat{n}_n}
G_0(\vec{r}_1,\vec{r}_n,E) \; . \end{equation}

For a system with a boundary ${\cal B}$ that is smooth everywhere, the
integrals in (\ref{z5}) can be evaluated in stationary phase
approximation. In this way $\mbox{Tr}\, \hat{Q}^n$ is expressed in
terms of a sum of contributions arising from periodic orbits with $n$
specular reflections. Inserting this approximation into Eq.\
(\ref{z4}) yields Gutzwiller's trace formula as shown, for example, in
\ref{Compo}.

The standard approach described above is not convenient for deriving
contributions of diffractive orbits, since the diffractive effects of
the corners are hidden in this formulation. Instead, we use a
modification of the boundary element method by formulating it in terms
of a Green function accounting for the diffractive effects of a
corner.

We restrict to the consideration of diffractive orbits with a single
corner point which are not influenced by the other corners of the
billiard. To obtain the contribution of such an orbit to the trace
formula it suffices to include the diffractive effect of only one
corner. We further restrict in this section to corners in which the
limit of the curvature of the boundary is zero when the corner is
approached from either side. Modifications caused by non-vanishing
curvature are discussed in Appendix D.

Let us first consider a simple billiard system which is bounded by a
wedge of angle $\gamma$ and an additional smooth curve $C$ which
connects the two sides of the wedge (such as represented in Fig.\ 6
for instance). One can derive an integral equation in terms of the
Green function $G_\gamma$ of the infinite wedge -- analogous to
equation (\ref{z1}) -- in which the integration is restricted to the
curve $C$. The oscillatory part of the spectral density is then again
given by (\ref{z4}) where the trace of $\hat{Q}^n$ now has the form

\begin{equation} \label{z6}
\mbox{Tr}\, \hat{Q}^n (k) = (-2)^n \int_C \! d s_1 \dots d s_n \,
\partial_{\hat{n}_1} \, G_\gamma(\vec{r}_2,\vec{r}_1,E) \,
\partial_{\hat{n}_2} \, G_\gamma(\vec{r}_3,\vec{r}_2,E) \dots
\partial_{\hat{n}_n} \, G_\gamma(\vec{r}_1,\vec{r}_n,E) \; .
\end{equation}

Note that similar techniques have been used in Ref.\ \cite{Pri95} for
deriving diffractive contributions in the Sinai billiard and in Ref.\
\cite{Pis96} for reformulating Fredholm's theory in the case of
triangles.

The Green function $G_\gamma$ can be split into a geometrical and a
diffractive part as has been done in section 2

\begin{equation} \label{z7}
G_\gamma(\vec{r},\vec{r}\,',E) = G_{geo}(\vec{r},\vec{r}\,',E) +
G_{diff}(\vec{r},\vec{r}\,',E) \; ,
\end{equation}

\noindent where the diffractive Green function is given by expressions
(\ref{e4a},\ref{e9},\ref{e10}). Inserting Eq.\ (\ref{z7}) into
(\ref{z6}) results in $2^n$ integrals. The stationary points of these
integrals correspond to periodic and diffractive orbits of the
billiard system (and possibly also to ghost orbits as in the case of
billiards without corners), and the number of points in a corner of a
diffractive orbit is determined by the number of diffractive parts
$G_{diff}$ appearing in the integral. Since we restrict to orbits with
one point in a corner we can replace $(n-1)$ of the Green functions in
(\ref{z6}) by their geometrical part. This can be done in $n$ ways
which cancels the factor $1/n$ in (\ref{z4}). Then the contribution to
the level density from orbits with $n$ reflections on the boundary $C$
and one point in a corner are contained in

\begin{equation} \label{z8}
\tilde{d}^{(n)}_1(k) = \frac{(-2)^n}{\pi} \Im \frac{d}{d k} \int_C \!
d s_1 \dots d s_n \, \partial_{\hat{n}_1} \,
G_\gamma(\vec{r}_2,\vec{r}_1,E) \partial_{\hat{n}_2} \,
G_{geo}(\vec{r}_3,\vec{r}_2,E) \dots \partial_{\hat{n}_n} \,
G_{geo}(\vec{r}_1,\vec{r}_n,E) \; ,
\end{equation}

\noindent an example is given in Fig.\ 5.
\begin{figure}[thb]
\begin{center}
\mbox{\epsfig{file=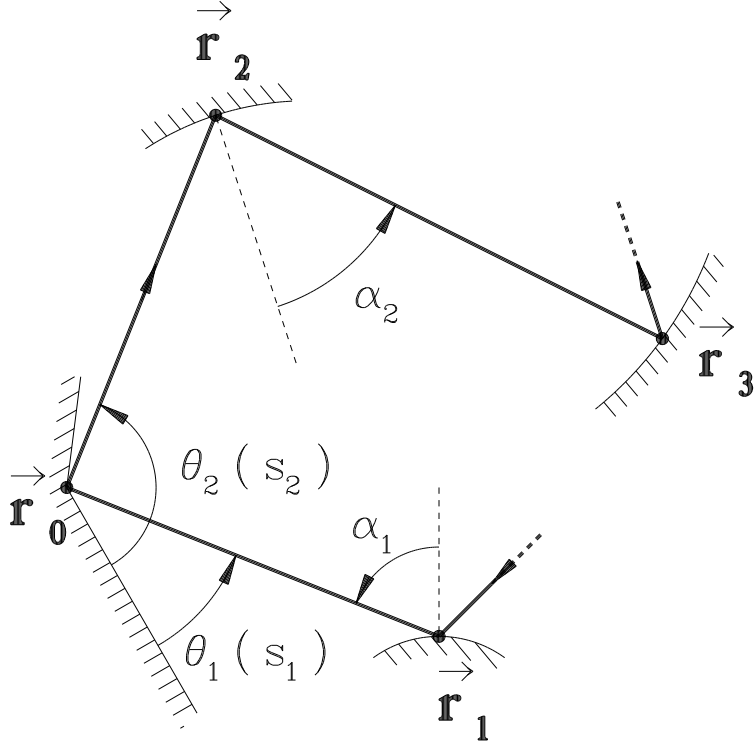,width=10cm,
bbllx=89pt, bblly=159pt, bburx=504pt, bbury=575pt,clip=}}
\end{center}
\caption{Typical path contributing to (\protect\ref{z8}). 
A precise definition of the angles $\alpha_i$ can be found
in Appendix B.}
\end{figure}

	Let us discuss Eq.\ (\ref{z8}) in more detail. The
contribution of a diffractive orbit is obtained by evaluating the
integrals in the vicinity of the stationary points, i.\,e.\ in the
vicinity of the points of specular reflection of the orbit on the
boundary. If one approximates $G_\gamma$ in the framework of GTD this
results in Eq.\ (\ref{z0}) (with $p=1$) for the contribution of the
diffractive orbit. In the following we will improve on this method by
using a uniform approximation for the Green function $G_\gamma$. In
both cases however, only local information about the reflection points
and the corner enters the approximation. It is then obvious how the
expression (\ref{z8}) has to be modified in order to derive the
semiclassical (or uniform) contributions to $d(k)$ for more
complicated diffractive orbits in billiards with several corners: for
every straight part between two reflection points a free Green
function has to be included, and for every part of the trajectory
which hits a corner between two reflections a Green function for an
infinite wedge with the same angle. The reason why $G_{geo}$ is
appearing in Eq.\ (\ref{z8}) and not $G_0$ is that in the above
formulation we consider only reflections on the part $C$ of the
boundary, and $G_{geo}$ takes care of the reflections on the wedge
part. Hence the total number of specular reflections on $\partial
{\cal B}$ in Eq.\ (\ref{z8}) may be greater than $n$.

We continue now with the further evaluation of Eq.\ (\ref{z8}) which
we perform in the case $n \neq 1$. The calculations for $n=1$ can be
treated by identical methods and yield the same final result.

In (\ref{z8}), $(n-2)$ boundary integrals can be evaluated by applying
the composition law (\ref{U1}) for Green functions which is derived in
Appendix B:

\begin{equation} \label{z9}
\partial_{\hat{n}_2} G_{sc} (\vec{r}_1,\vec{r}_2,E) \approx
(-2)^{(n-2)} \int_C \! d s_3 \dots d s_n \, \partial_{\hat{n}_2} \,
G_{geo}(\vec{r}_3,\vec{r}_2,E) \dots \partial_{\hat{n}_n} \,
G_{geo}(\vec{r}_1,\vec{r}_n,E) \; ,
\end{equation}

\noindent and consequently

\begin{equation} \label{z10}
\tilde{d}^{(n)}_1(k) = \frac{(-2)^2}{\pi} \Im \frac{d}{d k} \int_C \!
d s_1 \, d s_2 \, \partial_{\hat{n}_1} \,
G_\gamma(\vec{r}_2,\vec{r}_1,E) \partial_{\hat{n}_2} \, G_{sc}
(\vec{r}_1,\vec{r}_2,E) \; .
\end{equation}

	Here $G_{sc}$ is the contribution to the semiclassical Green
function from trajectories with $(n-2)$ reflections on the boundary
curve $C$ (and possibly further reflections on the wedge part of the
boun\-da\-ry):

\begin{equation} \label{z11}
G_{sc} (\vec{r}_1,\vec{r}_2,E) = \sum_{\xi} \frac{1}{\sqrt{8 \pi k
|m_{12}|}} \exp\{ i k l - i \frac{\pi}{2} \tilde{\nu} - i \frac{3
\pi}{4} \} \; ,
\end{equation}

\noindent where $m$ is the stability matrix (see \ref{Compo}) and $l$
the length of the classical orbit going from $\vec{r}_2$ to
$\vec{r}_1$.  $\tilde{\nu}$ is the number of conjugate points plus
twice the number of specular reflections on the boundary. We use here
$\tilde{\nu}$ and lower case letters for $m$ and $l$ in order to
distinguish these quantities from those of the whole diffractive
orbit. The normal derivative of the Green function is given in leading
order by

\begin{equation} \label{z12}
\partial_{\hat{n}_2} \, G_{sc} (\vec{r}_1,\vec{r}_2,E) \approx i k
\cos \alpha_2 \, G_{sc} (\vec{r}_1,\vec{r}_2,E) \; ,
\end{equation}

\noindent where $\alpha_2$ is the outgoing reflection angle at
$\vec{r}_2$ (see Fig.\ 5).

In the following, we will consider the contributions of the
geometrical and diffractive parts $G_{geo}$ and $G_{diff}$ to the
Green function $G_\gamma$ in Eq.\ (\ref{z10}) separately.  As
discussed above, the geometrical part will yield the contributions of
periodic orbits.  The reason why it has to be included also for the
derivation of the contributions of diffractive orbits is that both
$G_{diff}$ and $G_{geo}$ are discontinuous at the optical boundary
(see the discussion in Sec.\ 2). For that reason the boundary
contribution of $G_{geo}$ which arises from this discontinuity has to
be included in order to cancel the analogous contribution of
$G_{diff}$.

\subsection{The diffractive contribution}

	From (\ref{e4a},\ref{e3c},\ref{e9},\ref{e10}) the diffractive
part of the Green function $G_\gamma$ can be approximated by

\begin{equation} \label{z13}
G_{diff} (\vec{r}_2,\vec{r}_1,E) \approx \sum_{\sigma, \eta = \pm 1}
\sigma \eta \sqrt{\frac{2}{\pi k}} \frac{\displaystyle \mbox{\Large
e}^{-\displaystyle i\pi /4}}{16 \gamma} \int_{- i \infty}^{i \infty}
\! dz \, \frac{\exp \{ i k \sqrt{r_1^2 + r_2^2 - 2 r_1 r_2 \cos z} \}
}{ (r_1^2 + r_2^2 - 2 r_1 r_2 \cos z)^{1/4} \, \tan \left(
\frac{\displaystyle z + \phi_\sigma + \eta \pi}{\displaystyle 2 N}
\right)} \; ,
\end{equation}

\noindent where $r_1$ (resp.\ $r_2$) is the distance from $\vec{r}_1$
(resp.\ $\vec{r}_2$) to the diffractive point, and $\phi_\sigma =
\theta_1 - \sigma\theta_2$ (see Fig.\ 5). Similarly to (\ref{z12}),
the normal derivative $\partial_{\hat{n}_1}$ yields a factor $i k \cos
\alpha_1$. We insert (\ref{z13}) and one contribution $\xi$ from
(\ref{z11}) into (\ref{z10}) and consider the contribution from the
vicinity of a stationary point which is chosen as origin of the
$s$-variables. The main contribution to the $z$-integral comes from
values near $z=0$ and the exponent is expanded in $z$ up to second
order

\begin{eqnarray} \label{z14}
d_{\xi,diff} (k) &\approx& \frac{(-2)^2}{\pi} \Im \frac{d}{dk}
\sum_{\sigma, \eta = \pm 1} \sigma \eta \sqrt{\frac{2}{\pi k}}
\frac{\mbox{\Large e}^{\displaystyle -i \pi /4}}{16 \gamma}
\frac{\exp\{- i \frac{\pi}{2} \tilde{\nu} - i \frac{3 \pi}{4} \}
}{\sqrt{8 \pi k |m_{12}|}} (i k)^2 \cos \alpha_1 \cos \alpha_2
\nonumber \\ && \int_C \! d s_1 \, d s_2 \, \int_{- i \infty}^{i
\infty} \! dz \, \frac{\exp \{ i k (l(s_1,s_2) + r_1(s_1) + r_2(s_2) -
\frac{\displaystyle r_1 r_2}{\displaystyle 2(r_1+r_2)} z^2 ) \} }
{\displaystyle \sqrt{r_1 + r_2} \, \tan \left( \frac{\displaystyle z +
\phi_\sigma + \eta \pi}{\displaystyle 2 N} \right) } \; ,
\end{eqnarray}

\noindent where the index $\xi$ labels the diffractive orbit.  A
stationary phase approximation of all integrals would yield the
contribution of the diffractive orbit in the GTD approximation. This
approximation diverges at an optical boundary. In order to obtain a
finite uniform approximation the effect of the nearest pole to $z=0$
has to be included. We treat this pole again by the method of Pauli

\begin{eqnarray} \label{z15}
\frac{1}{\tan \left( \frac{\displaystyle z + \phi_\sigma + \eta
\pi}{\displaystyle 2 N} \right) } & = & \frac{1} {\tan \left(
\frac{\displaystyle z + \phi_\sigma + \eta \pi}{\displaystyle 2 N}
\right) } \; \frac{a_{\sigma,\eta} + \eta \sqrt{2} \sin
\frac{\displaystyle z + \Delta \phi_\sigma}{\displaystyle 2}} {
a_{\sigma,\eta} + \eta \sqrt{2} \sin \frac{\displaystyle z + \Delta
\phi_\sigma}{\displaystyle 2}} \nonumber \\ & \approx & \frac{1} {\tan
\left( \frac{\displaystyle\phi_{\sigma,0} + \eta \pi}{\displaystyle2
N}\right) } \; \frac{a_{\sigma,\eta}}{a_{\sigma,\eta} + \eta \,
\frac{\displaystyle z + \Delta \phi_\sigma}{\displaystyle\sqrt{2}} }
\;\; ,
\end{eqnarray}

\noindent where $\phi_{\sigma,0}$ is the value of $\phi_\sigma$ at the
stationary point, $\Delta \phi_\sigma = \phi_\sigma - \phi_{\sigma,0}$
and $a_{\sigma,\eta}$ is evaluated using (\ref{e12p}) at the
stationary point.  Inserting (\ref{z15}) into (\ref{z14}) we obtain

\begin{eqnarray} \label{z16}
d_{\xi,diff} (k) &\approx& \Im \frac{d}{dk} \sum_{\sigma, \eta = \pm
1} \sigma \eta \frac{k \cos \alpha_1 \cos \alpha_2 \exp\{- i
\frac{\pi}{2} \tilde{\nu} \} }{8 \pi^2 \gamma \sqrt{(r_1 + r_2)
|m_{12}|}} \frac{a_{\sigma,\eta}} {\tan \left( \frac{\displaystyle
\phi_{\sigma,0} + \eta \pi}{\displaystyle 2 N} \right) } \nonumber \\
&& \int_{-\infty}^\infty \! d s_1 \, d s_2 \, \int_{- i \infty}^{i
\infty} \! dz \, \frac{\exp \{ i k (l(s_1,s_2) + r_1(s_1) + r_2(s_2) -
\frac{\displaystyle r_1 r_2}{\displaystyle 2(r_1+r_2)} z^2 ) \} }{
a_{\sigma,\eta} + \eta (z + \Delta \phi_\sigma) / \sqrt{2}} \; .
\end{eqnarray}

	The quantities $l(s_1,s_2)$, $r_1(s_1)$ and $r_2(s_2)$ are now
 expanded up to second order in $s_1$ and $s_2$. The expansion
 coefficients can be obtained from (\ref{U8}). Furthermore we expand
 $\Delta \phi_\sigma$ up to first order in $s_1$ and $s_2$: $\Delta
 \phi_\sigma (s_1,s_2) \approx s_1 \cos \alpha_1/r_1 - \sigma s_2 \cos
 \alpha_2/r_2$.  After a substitution

\begin{equation} \label{z18}
s_1 \rightarrow - \eta \frac{\sqrt{2}}{\cos \alpha_1} s_1 \; , \; \;
\; s_2 \rightarrow - \eta \frac{\sqrt{2}}{\cos \alpha_2} s_2 \; , \;
\; \; z \rightarrow - \eta \sqrt{2} z \; ,
\end{equation}

\noindent we obtain the following expression

\begin{eqnarray} \label{z19}
d_{\xi,diff} (k) &\approx& - \Im \frac{d}{dk} \sum_{\sigma, \eta = \pm
1} \sigma \eta \frac{k \exp\{- i \frac{\pi}{2} \tilde{\nu} \} }{\pi^2
\gamma \sqrt{8 (r_1 + r_2) |m_{12}|}} \frac{a_{\sigma,\eta}} {\tan
\left(\displaystyle\frac{\phi_{\sigma,0} + \eta \pi}{\displaystyle 2
N} \right) } \nonumber \\ && \int_{-\infty}^\infty \! d s_1 \, d s_2
\, \int_{- i \infty}^{i \infty} \! dz \, \frac{\exp \{ i k (l + r_1 +
r_2 + a' s_1^2 + b' s_2^2 + \frac{\displaystyle 2}{\displaystyle
m_{12}} s_1 s_2 - c z^2) \} } {z + \frac{\displaystyle
s_1}{\displaystyle r_1} - \sigma \frac{\displaystyle
s_2}{\displaystyle r_2} - a_{\sigma,\eta}} \; ,
\end{eqnarray}

\noindent where

\begin{equation} \label{z20}
a'=\frac{m_{11}}{m_{12}}-\frac{2}{R_1 \cos \alpha_1}+\frac{1}{r_1} \;
, \; \; \; b'=\frac{m_{22}}{m_{12}}-\frac{2}{R_2 \cos
\alpha_2}+\frac{1}{r_2} \; , \; \; \; c = \frac{r_1 r_2}{r_1 + r_2} \;
.
\end{equation}

	Here the quantities $r_1$, $r_2$ and $l$ without argument
denote the values at the stationary point.  The derivative with
respect to $k$ in (\ref{z19}) yields in leading order a factor $i L$
where $L=r_1+r_2+l$ is the length of the diffractive orbit.  In the
next step we simplify the integrals by applying a transformation of
the $s$ variables such that the denominator of the integrand depends
only on one of the new $s$ variables

\begin{equation} \label{z21}
s = \frac{s_1}{r_1} - \sigma \frac{s_2}{r_2} \; , \; \; \; s' = d \,
r_2 s_1 + (1 - \sigma \, d) \, r_1 s_2 \; .
\end{equation}

	The form of $s'$ is chosen such that the Jacobian of the
transformation is one and the value of $d$ is determined by the
requirement that the exponent in the integrand has no mixed quadratic
term $s's$. The evaluations are done with Maple and result in

\begin{eqnarray} \label{z22}
d_{\xi,diff} (k) &\approx& - \Re \sum_{\sigma, \eta = \pm 1} \sigma
\eta \frac{k L \exp\{i k L - i \frac{\pi}{2} \tilde{\nu} \} }{\pi^2
\gamma \sqrt{8 (r_1 + r_2) |m_{12}|}} \frac{a_{\sigma,\eta}} {\tan
\left( \frac{\displaystyle\phi_{\sigma,0} + \eta \pi}{\displaystyle 2
N} \right) } \nonumber \\ && \int_{-\infty}^\infty \! ds \, d s' \,
\int_{- i \infty}^{i \infty} \! dz \, \frac{\exp \{ i k (a s^2 + b
s'^2 - c z^2) \} }{ z + s - a_{\sigma,\eta}} \; ,
\end{eqnarray}

\noindent where

\begin{eqnarray} \label{z23}
a &=& \frac{M_{12} c}{M_{12} - c(\mbox{Tr}\, M - \sigma 2)} \; , \; \;
\; b = \frac{M_{12} - c(\mbox{Tr}\, M - \sigma 2)}{m_{12} r_1 r_2 c}
\; , \\ \nonumber d &=& \frac{c(\sigma r_1 R_1 \cos \alpha_1 m_{11} -
2 \sigma r_1 m_{12} + \sigma R_1 \cos \alpha_1 m_{12} + r_2 R_1 \cos
\alpha_1)}{ r_2 R_1 \cos \alpha_1 [M_{12} - c(\mbox{Tr}\, M - \sigma
2)]} \; ,
\end{eqnarray}

	$M$ is the stability matrix of the diffractive orbit, i.\,e.\
the stability matrix of the classical trajectory starting from the
corner going through $\vec{r}_2$ and $\vec{r}_1$ and back to the
corner. There is a relation between $\tilde{\nu}$, the signs of $a$
and $b$ and the Maslov index $\nu$ of the diffractive orbit

\begin{equation} \label{z24}
\exp \{ - i \frac{\pi}{2} \tilde{\nu} + i \frac{\pi}{4} \sigma_a + i
\frac{\pi}{4} \sigma_b \} = i \exp \{ - i \frac{\pi}{2} \nu \} \; ,
\end{equation}

\noindent where $\sigma_a = \mbox{sign} (a)$, $\sigma_b = \mbox{sign}
(b)$.  Hence $\nu$ is equal to $\tilde{\nu}$ plus the number of
negative signs of $a$ and $b$ (modulo 4). This can be seen, for
example, by evaluating the $s_1$- and $s_2$-integrals in Eq.\
(\ref{z19}) by the stationary phase method. Then from the composition
law of Green functions the Maslov index $\nu$ of the whole diffractive
orbit is obtained by successive applications of (\ref{U13}).  Since a
stationary phase approximation after the transformation (\ref{z21})
has to yield the same result, relation (\ref{z24}) follows
immediately.

	The integral over $s'$ can now be evaluated, and the double
integral over $s$ and $z$ is calculated in \ref{Diffint} (Eq.\
(\ref{B10})).  The result is

\begin{eqnarray} \label{z25}
d_{\xi,diff} (k) &\approx& - \Re \sum_{\sigma, \eta = \pm 1} \sigma
\eta \tau_\sigma \frac{L \exp\{i k L - i \frac{\pi}{2} \tilde{\nu} \}
}{\gamma \sqrt{8 (r_1 + r_2) |m_{12} b (a-c)|}}
\frac{|a_{\sigma,\eta}|} {\tan \left(
\frac{\displaystyle\phi_{\sigma,0} + \eta \pi}{\displaystyle 2 N}
\right) } \mbox{\Large e}^{i\pi (1 + \sigma_a + \sigma_b +
\tau_\sigma)/4} \nonumber \\ && \exp\{ - \frac{i k a c
a_{\sigma,\eta}^2}{a-c} \} \, \left[ \mbox{erfc} \{ |a
a_{\sigma,\eta}| \sqrt{\frac{k}{i(a-c)}} \} - \mbox{erfc} \{
|a_{\sigma,\eta}| \sqrt{\frac{kac}{i(a-c)}} \} \right] \; ,
\end{eqnarray}

\noindent where $\tau_\sigma = \mbox{sign} (ac/(a-c)) = \mbox{sign}
(M_{12}/(\mbox{Tr}\, M - \sigma 2))$.  As will be seen in the
following section, the first error function in (\ref{z25}) is the
contribution from the discontinuity of $G_{diff}$ which is cancelled
by the corresponding contribution from $G_{geo}$.

\subsection{The geometrical contribution}

For a given $\sigma$ and $\eta$ the geometrical orbit that corresponds
to the nearest pole arises when $\phi_\sigma - 2 n_{\sigma,\eta}
\gamma = - \eta \pi$ and it exists if $\eta(2 n_{\sigma,\eta} \gamma -
\phi_\sigma) < \pi$.  This can be re-expressed in the form

\begin{equation} \label{z26}
a_{\sigma,\eta} = \sqrt{2} \cos(\frac{\phi_{\sigma,0} - 2
n_{\sigma,\eta} \gamma}{2}) > - \sqrt{2} \sin \frac{\eta \Delta
\phi_\sigma}{2} \approx - \frac{\eta \Delta \phi_\sigma}{\sqrt{2}} \;
.
\end{equation}

	In the following we derive the contribution from the
discontinuity of $G_{geo}$. For that purpose we apply exactly the same
approximations to $G_{geo}$ that were used for $G_{diff}$. This is
done by writing $G_{geo}$ in the form

\begin{eqnarray} \label{z27}
G_{geo} (\vec{r}_2,\vec{r}_1,E) &\approx& - \! \sum_{\sigma, \eta =
\pm 1} \! \sigma \Theta(A) \sqrt{\frac{2}{\pi k}} \,
\frac{\mbox{\Large e}^{\displaystyle i\pi /4}}{4}\; \frac{\exp \{ i k
\sqrt{r_1^2 + r_2^2 - 2 r_1 r_2 \cos (\phi_\sigma - 2 n_{\sigma,\eta}
\gamma)} \} }{ (r_1^2 + r_2^2 - 2 r_1 r_2 \cos (\phi_\sigma - 2
n_{\sigma,\eta} \gamma))^{1/4}} \\ \nonumber &=& - \! \sum_{\sigma,
\eta = \pm 1} \! \sigma \Theta(A) \sqrt{\frac{2}{\pi k}} \,
\frac{\mbox{\Large e}^{\displaystyle -i\pi /4}}{16 \gamma} \oint \! dz
\, \frac{\exp \{ i k \sqrt{r_1^2 + r_2^2 - 2 r_1 r_2 \cos z} \} }{
(r_1^2 + r_2^2 - 2 r_1 r_2 \cos z)^{1/4} \, \tan \left(
\frac{\displaystyle z + \phi_\sigma + \eta \pi}{\displaystyle 2 N}
\right) } \; ,
\end{eqnarray}

\noindent where $A=a_{\sigma,\eta} + \sqrt{2} \sin (\eta \Delta
\phi_\sigma/2)$.  The integration contour of the $z$-integral
encircles the nearest pole to $z=0$ counter-clockwise. The expression
(\ref{z27}) differs from Eq.\ (\ref{z13}) only by a factor $(-\eta)$,
the $\Theta$-function and the integration contour.  We repeat now all
the steps from Eq.\ (\ref{z14}) to (\ref{z22}). The only difference is
an multiplicative factor $(- \eta)$ which results from the
substitution $z \rightarrow - \eta \sqrt{2} z$ (it didn't appear
previously because of the different integration contour).  We arrive
at an expression corresponding to Eq.\ (\ref{z22})

\begin{eqnarray} \label{z28}
&& - \Re \sum_{\sigma, \eta = \pm 1} \sigma \eta \frac{k L \exp\{i k L
- i \frac{\pi}{2} \tilde{\nu} \} }{\pi^2 \gamma \sqrt{8 (r_1 + r_2)
|m_{12}|}} \; \frac{a_{\sigma,\eta}} {\tan \left(
\frac{\displaystyle\phi_{\sigma,0} + \eta \pi}{\displaystyle 2 N}
\right) } \nonumber \\ && \int_{-\infty}^\infty \! ds \, d s' \, \oint
\! dz \, \Theta(a_{\sigma,\eta} - s) \frac{\exp \{ i k (a s^2 + b s'^2
- c z^2) \} }{ z + s - a_{\sigma,\eta}} \; .
\end{eqnarray}

	The triple integral is denoted by $I$. The integrals over $s'$
and $z$ can now be evaluated and result in

\begin{eqnarray} \label{z29}
I &=& 2 \pi i \, \sqrt{\frac{\pi i}{k b}}
\int_{-\infty}^{a_{\sigma,\eta}} \! ds \, \exp \{i k a s^2 - i k c
(s-a_{\sigma,\eta})^2 \} \nonumber \\ &=& 2 \pi i \, \sqrt{\frac{\pi
i}{k b}} \int_{-\infty}^0 \! ds \, \exp \{ i k (a-c) (s + \frac{a
a_{\sigma,\eta}}{a-c} )^2 - i \frac{k a c a_{\sigma,\eta}^2}{a-c} \}
\; .
\end{eqnarray}

	We are interested only in the boundary contribution of the
geometrical part. The expression (\ref{z28}) contains in general also
contributions from periodic orbits. This is the case if the
integration range in (\ref{z29}) contains a stationary point, i.\,e.\
if $a a_{\sigma,\eta}/(a-c)$ is positive.  Then the stationary point
contribution has to be subtracted which corresponds to a subtraction
of the integral from $-\infty$ to $\infty$. We obtain for the boundary
contribution

\begin{eqnarray} \label{z30}
I' &=& - \mbox{sign}(a_{\sigma,\eta}) \, \tau_\sigma 2 \pi i
\sqrt{\frac{\pi i}{k b}} \int_0^\infty \! ds \exp \{ i k (a-c) (s +
|\frac{a a_{\sigma,\eta}}{a-c}| )^2 - i \frac{k a c
a_{\sigma,\eta}^2}{a-c} \} \nonumber \\ &=& -
\mbox{sign}(a_{\sigma,\eta}) \, \tau_\sigma \frac{\pi^2}{k}
\frac{1}{\sqrt{|b (a-c)|}} \mbox{\Large e}^{i \frac{\pi}{4}
(1+\sigma_a+\sigma_b+\tau_\sigma)} \exp \{- i \frac{k a c
a_{\sigma,\eta}^2}{a-c} \} \mbox{erfc} \{ |a a_{\sigma,\eta}|
\sqrt{\frac{k}{i(a-c)}} \} \; ,
\end{eqnarray}

	where $\tau_\sigma = \mbox{sign} (ac/(a-c))$, as before.
Substituting $I'$ for the triple integral in (\ref{z28}) yields

\begin{eqnarray} \label{z31}
d_{\xi,geo} (k) &\approx& \Re \sum_{\sigma, \eta = \pm 1} \sigma \eta
\tau_\sigma \frac{L \exp\{i k L - i \frac{\pi}{2} \tilde{\nu} \}
}{\gamma \sqrt{8 (r_1+r_2)|m_{12} b (a-c)|}} \frac{|a_{\sigma,\eta}|}
{\tan \left( \frac{\displaystyle\phi_{\sigma,0} + \eta
\pi}{\displaystyle 2 N} \right) } \nonumber \\ && \mbox{\Large e}^{i
\frac{\pi}{4} (1+\sigma_a+\sigma_b+\tau_\sigma)} \exp\{ - \frac{i k a
c a_{\sigma,\eta}^2}{a-c} \} \, \mbox{erfc} \{ |a a_{\sigma,\eta}|
\sqrt{\frac{k}{i(a-c)}} \} \; .
\end{eqnarray}

Comparison with Eq.\ (\ref{z25}) shows that this contribution exactly
cancels the first error function in (\ref{z25}).

\subsection{The joint contribution}

We now can write down the final formula of this section.  By using the
definitions of $a$, $b$ and $c$ the sum of (\ref{z25}) and (\ref{z31})
can be written in the form

\begin{eqnarray} \label{z32}
d_\xi (k) &\approx& - \Re \sum_{\sigma, \eta = \pm 1} \eta \tau_\sigma
\; \frac{L}{\pi} \; \frac{\exp\{i k L - i \frac{\pi}{2} \mu_\sigma \}
}{\sqrt{| \mbox{Tr}\, M - \sigma 2|}} \frac{|a_{\sigma,\eta}|}{2 N
\sqrt{2} \tan \left( \frac{\displaystyle \phi_\sigma + \eta
\pi}{\displaystyle 2 N} \right) } \nonumber \\ && \exp\{ - \frac{i k
a_{\sigma,\eta}^2 M_{12}}{\mbox{Tr}\, M - \sigma 2} \} \, \mbox{erfc}
\{ |a_{\sigma,\eta}| \sqrt{\frac{k M_{12}}{i(\mbox{Tr}\, M - \sigma
2)}} \} \; ,
\end{eqnarray}

\noindent
where we dropped the second index of $\phi_{\sigma,0}$ for simplicity
of notation. Furthermore, $\mu_\sigma = \nu + (1 - \sigma) +
\kappa_\sigma$, $\tau_\sigma=1-2\kappa_\sigma$, and $\kappa_\sigma$ is
defined as

\begin{equation} \label{z33}
\kappa_\sigma = \left\{ \begin{array}{rcl} 0 & \mbox{if} &
\frac{\displaystyle M_{12}}{\displaystyle \mbox{Tr}\, M - \sigma 2} >
0 \; , \\ & & \\ 1 & \mbox{if} & \frac{\displaystyle
M_{12}}{\displaystyle \mbox{Tr}\, M - \sigma 2} < 0 \; .
\end{array} \right.
\end{equation}

	We recall that $M$ in (\ref{z32},\ref{z33}) is the stability
matrix (at unit energy) of the classical trajectory starting and
ending at the corner point. $\nu$ is the number of conjugate points of
this trajectory, plus 2 times the number of specular reflections. The
definition of $\mu_\sigma$ in terms of $\nu$ and $\kappa_\sigma$ is
similar to the definition of the Maslov index of a periodic orbit in
terms of that of the Green function \cite{Cre90}.  The
$\sigma$-dependence is due to the fact that positive $\sigma$ values
are associated with geometrical orbits that are reflected an even
number of times near the corner, as an optical boundary is approached,
and negative $\sigma$ values with orbits with an odd number of
bounces. This is explained in more detail in Appendix D. In the
limiting case that the diffractive orbit becomes a periodic orbit, its
contribution comes only from one of the values of $\sigma$ (the other
cancels), and its stability matrix is $M$ or $-M$ depending on whether
the number of bounces of the orbit in the corner is even or odd (cf.
the discussion in 4.2). Thus $\mu_\sigma$ is identical to the Maslov
index of the periodic orbit in these limiting cases.

	In terms of the Fresnel integral $K$ formula (\ref{z32}) can
written in a slightly shorter form

\begin{equation} \label{z34}
d_\xi (k) \approx - \Re \sum_{\sigma, \eta = \pm 1} \eta \tau_\sigma
\; \frac{L}{\pi} \; \frac{\exp\{i k L - i \frac{\pi}{2} \mu_\sigma \}
}{\sqrt{| \mbox{Tr}\, M - \sigma 2|}} \frac{|a_{\sigma,\eta}|}{N
\sqrt{2} \tan \left( \frac{\displaystyle\phi_\sigma + \eta
\pi}{\displaystyle 2 N} \right) } K\left( |a_{\sigma,\eta}|
i^{\kappa_\sigma} \sqrt{\frac{k |M_{12}|}{|\mbox{Tr}\, M - \sigma 2|}}
\, \right) \; .
\end{equation}

	Equation (\ref{z34}) is the main result of this paper. It
gives a uniform approximation for the contribution of an isolated
diffractive orbit with a single corner point to the trace formula. For
completeness we recall several definitions: $\phi_{\sigma}=\theta_1 -
\sigma\theta_2$, where $\theta_1$ and $\theta_2$ are the incoming and
outgoing angles at the diffractive point (measured from the same edge,
with $\theta_1$ and $\theta_2 \in [0,\gamma]$) and $a_{\sigma,\eta}$
is defined by

\begin{equation}\label{z35}
a_{\sigma,\eta} = \sqrt{\displaystyle 2} \,
\cos({\displaystyle\phi_\sigma\over\displaystyle 2} -n_{\sigma,\eta}
\gamma) \quad\mbox{with}\quad n_{\sigma,\eta}=
\,\mbox{nint}\,[{\displaystyle\phi_\sigma +
\eta\,\pi\over\displaystyle 2\gamma}] \in \mbox{Z$\!\!$Z} \;
.\end{equation}

	Note finally that the modified Fresnel function of imaginary
argument (encountered when $\kappa_\sigma=1$) can be computed
numerically from (\ref{f1b}).

\section{Discussion of the result}

	In this section we discuss properties and the range of
validity of formula (\ref{z34}). As mentioned above, the derivation
has been done for a specific case (a wedge connected to a smooth
boundary), but it is more generally valid because it relies only on
local properties of the system near the considered diffractive orbit
(as usual in semiclassical approximations).  Hence it applies to
billiards of any shape provided the singularity of the boundary
corresponds locally to the intersection of two straight lines.
However, the present approach has to be refined if applied to curved
edges, we discuss this point in Appendix D. We also remind that
formula (\ref{z34}) is only valid for single diffraction. The same
formalism can in principle also be applied to diffractive orbits with
more than one diffractive point, the formulas become however
increasingly more complex. For example, in the case of double
diffraction, one has already 16 instead of 4 terms, and they involve
also double Fresnel integrals as can be inferred from the treatment of
diffraction at two wedges in \cite{Sch91}. The formulas can only be
simplified if the diffraction in some of the corners can be treated in
the GTD approximation.

Note also, that the factor $|\,\mbox{Tr}\, M - 2\sigma|^{-1/2}$ in
(\ref{z34}) diverges for a parabolic diffractive orbit (i.\,e.\ when
$\mbox{Tr}\, M=\pm 2$) and the present approach cannot be used in this
case. This is very similar to divergences in Gutzwiller's trace
formula due to non-isolated orbits.  For diffractive orbits, the fact
that $\mbox{Tr}\, M=\pm 2$ can have several reasons, for example the
diffractive orbits can appear in families as is the case in a circular
sector, or bifurcations of diffractive orbits can occur, or the
diffractive orbit can become a part of a family of periodic orbits
when the optical boundary is approached. The latter case can occur for
example in triangular billiards. In this case it is however often
possible to treat the divergent part (one of the $\sigma$-values) in
the GTD approximation if the diffractive orbit is well separated from
the torus of periodic orbits, and apply the uniform approximation only
to the non-divergent part as will be demonstrated in a numerical
example in section 5.

\subsection{The GTD limit}

	After these basic remarks we now study three simple limits of
Eq.  (\ref{z34}). The first one is the geometrical theory of
diffraction which is valid sufficiently far away from the optical
boundary. In this limit the argument of the $K$ function is large and
the function can be replaced by its leading asymptotic term in
(\ref{f2}). This immediately yields

\begin{equation}\label{dr1} d_\xi (k) \approx {\displaystyle 
L\over\pi} \, {\displaystyle {\cal D}
(\theta_1,\theta_2)\over\sqrt{\displaystyle 8\pi k|M_{12}|}} \, \cos
(k L- \nu\pi/2-3\pi/4) \; ,\end{equation}

\noindent which agrees with the general formula (\ref{z0}) in the case
 of one diffractive point. Analogous formulae have been derived and
 tested in \cite{Vat94,Pav95,Bru95}.  They have the advantage of
 allowing to treat general diffractive problems (other than wedge
 diffraction) and can easily be generalized to multiple diffraction
 (see (\ref{z0})). However they diverge on the optical boundary and
 (as shown in the examples 5.1 and 5.2 below) they are unable to
 describe the limit that a diffractive orbit is close to become a real
 trajectory.

\subsection{The limit $\gamma = \pi/p$}

	Let us now study the limit that the diffraction angle $\gamma$
 goes to $\pi/p$ ($p\in\mbox{I$\!$N}^*$). For these values of $\gamma$
 there is no diffraction since the corner can be treated by the method
 of images. As a consequence, the contributions of most diffractive
 orbits disappear, but there are also diffractive orbits which are
 replaced by periodic orbits which contribute to the level density
 according to Gutzwiller trace formula. Their contribution can be
 obtained from the diffractive contribution (\ref{z34}) in the limit
 $\gamma \rightarrow \pi/p$. The situation is actually slightly more
 complicated, since the diffractive contribution of these orbits for
 angles $\gamma=\pi/p+\epsilon$ is discontinuous at $\epsilon=0$ (it
 changes sign). The reason for this is that periodic orbits split from
 the diffractive orbit as $\epsilon$ goes through zero (for example as
 the billiard is deformed), which can be con\-si\-de\-red as a kind of
 bifurcation. As a consequence both diffractive and periodic orbit
 contributions are discontinuous at $\epsilon=0$, but their sum
 remains continuous. In order to discuss this in more detail we have
 to consider the cases of odd and even $p$ separately.

\

$\bullet$ Case $\gamma = \frac{\pi}{2 p} + \epsilon$. In the limit
$\epsilon=0$ the contributions from the two $\eta$-values cancel for
$\sigma = -1$. The same occurs for $\sigma = +1$, except if $\theta_2
= \theta_1$. If this condition is fulfilled one obtains

\begin{equation}\label{dr2} d_\xi (k)\to\,\mbox{sign}\,(\epsilon) \, 
\tau_{(+)} \, d_{po}(k) \; ,\end{equation}

\noindent where

\begin{equation}\label{dr3} d_{po}(k) = {\displaystyle L\over\pi} \; 
{\displaystyle \cos(k L-\mu_{(+)}\pi/2) \over\displaystyle
|\,\mbox{Tr}\, M - 2|^{1/2}} \qquad\qquad (\;\mbox{when}\quad
\gamma={\displaystyle\pi\over\displaystyle 2p}) \; .\end{equation}

	The discontinuity in Eq.\ (\ref{dr2}) at $\epsilon = 0$ is
directly related to the appearance of new periodic orbits. This can be
seen from the discussion in section 3.2: in Eq.\ (\ref{z29}) one has
contributions of periodic orbits in the vicinity of the diffractive
orbit if $a_{\sigma,\eta} \tau_\sigma > 0$, and the periodic orbits
coincide with the diffractive orbit when $a_{\sigma,\eta} = 0$. For
the considered case the above inequality is equivalent to
$-\mbox{sign}(\epsilon) \tau_{+} > 0$. Hence when $\epsilon$ goes
through zero, two periodic orbits appear (or disappear), one for each
value of $\eta$, assuring the continuity of the sum of contributions
at $\epsilon = 0$.

$\bullet$ Case $\gamma=\pi/(2p+1)+\epsilon$. Now the two contributions
 to $\sigma = +1$ cancel as $\epsilon \rightarrow 0$, and for $\sigma
 = -1$ there is only a contribution if $\theta_2 = \gamma -
 \theta_1$. This contribution is of the form

\begin{equation}\label{dr4} d_\xi (k)\to\,\mbox{sign}\,(\epsilon) \, 
\tau_{(-)} \, d_{po}(k) \; ,\end{equation}

\noindent and the periodic orbit contribution now is given by

\begin{equation}\label{dr5} d_{po}(k) = {\displaystyle L\over\pi} \;
{\displaystyle \cos(k L-\mu_{(-)}\pi/2) \over\displaystyle
|\,\mbox{Tr}\, M + 2|^{1/2}} \qquad\qquad (\;\mbox{when}\quad
\gamma={\displaystyle\pi\over\displaystyle 2p+1}) \; .\end{equation}

	Comparing with (\ref{dr3}), the reason for the change of sign
of $M$ is the odd number of classical reflections on the vertex in the
case $\gamma=\pi/(2p+1)$. Generally, the stability matrix $M$ of the
closed trajectory in (\ref{z34}) becomes equal to plus (resp. minus)
the monodromy matrix of the periodic orbit when $\gamma$ is $\pi/(2p)$
(resp.\ $\pi/(2p+1)$).  The explanation of the discontinuity of Eq.\
(\ref{dr4}) is the same as above with the only difference that the
condition for the existence of neighbouring periodic orbits can now be
expressed by $-\mbox{sign}(\epsilon) \tau_{-} > 0$.

In billiards with corners one has therefore a new kind of bifurcation:
 the continuity of wave mechanics (in the semiclassical approximation)
 is not enforced by complex trajectories but by diffractive orbits.
 This effect will be demonstrated in the examples below.

\subsection{In the vicinity of an optical boundary}

The case that a diffractive orbit lies on an optical boundary, or
crosses an optical boundary when the billiard is deformed, is very
similar to the case $\gamma \rightarrow \pi/p$. Again the diffractive
orbit contributes on the optical boundary at the same order of $k$ as
a periodic orbit, but now only with half the amplitude of a periodic
orbit. The diffractive contribution is again discontinuous since it
changes sign as an optical boundary is crossed, and the reason for
this is that a new periodic orbit arises which bifurcates from the
diffractive orbit. More specifically, let us consider the case that
for a given value of $\sigma$ and $\eta$ one has $\phi_\sigma - 2
n_{\sigma,\eta} \gamma + \eta \pi = \epsilon$ where $\epsilon$ is
small. In the limit $\epsilon \rightarrow 0$ the contribution from
these values of $\sigma$ and $\eta$ to the spectral density is given
by

\begin{equation}\label{dr6} - \frac{1}{2} \, \eta \tau_\sigma\,
\mbox{sign}\,(\epsilon) \, d_{po}(k) \qquad\mbox{where} \qquad
d_{po}(k) = {\displaystyle L\over\pi} \; {\displaystyle \cos(k
L-\mu_\sigma\pi/2) \over\displaystyle \sqrt{|\,\mbox{Tr}\, M - 2
\sigma|}} \; .\end{equation}

and one can verify that the discontinuity of Eq.\ (\ref{dr6}) is due
to a neighbouring periodic orbit which coincides with the diffractive
orbit at $\epsilon = 0$. As above, the condition for the existence of
the periodic orbit is $a_{\sigma,\eta} \tau_\sigma > 0$ which now is
equivalent to $\eta \epsilon \tau_\sigma > 0$.

\section{Some examples}

\begin{figure}[thbp]
\begin{center}
\mbox{\epsfig{file=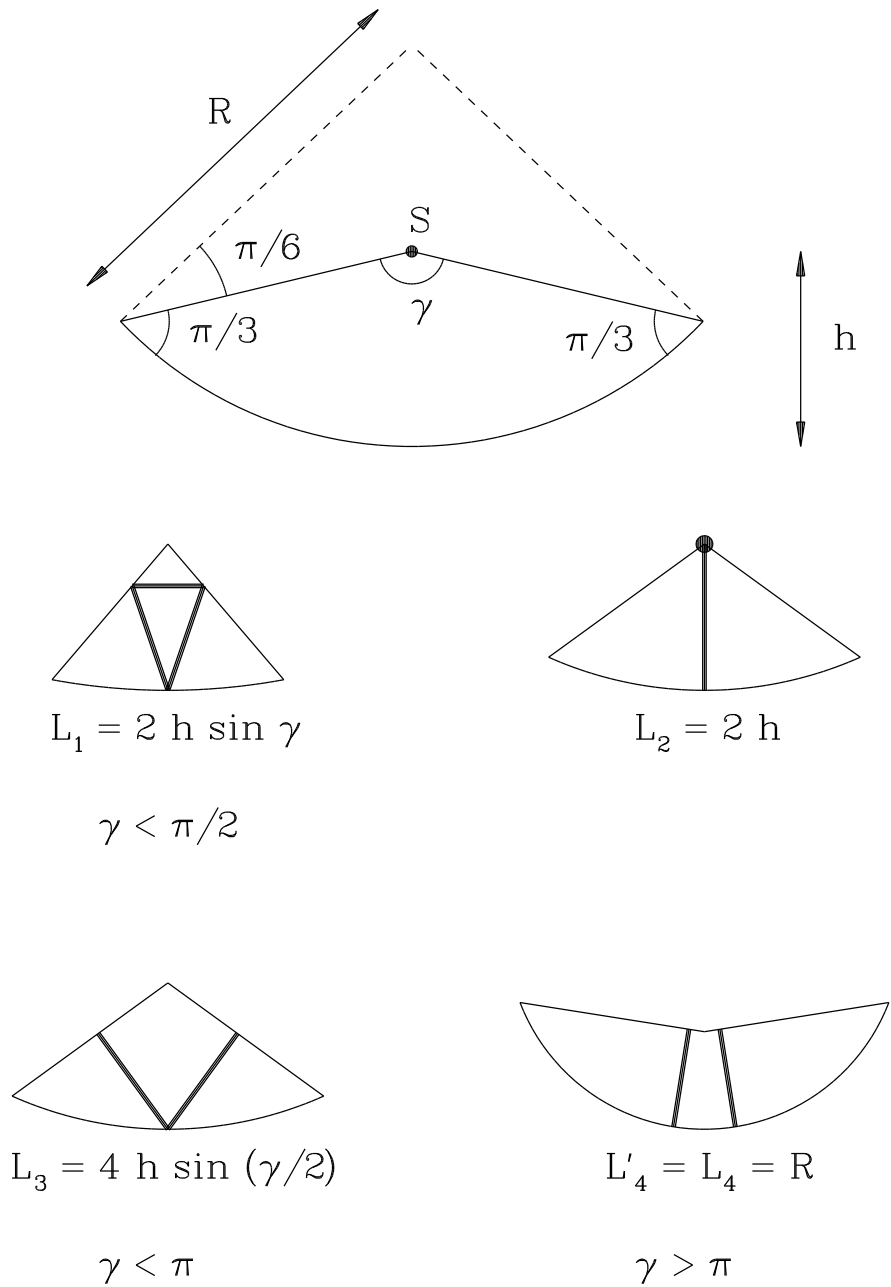,width=12cm,
bbllx=111pt, bblly=89pt, bburx=470pt, bbury=608pt,clip=}}
\end{center}
\caption{Shortest periodic and diffractive orbits in the
``rounded triangular billiard" studied in Sec.\ 5. For the
diffractive orbit, the diffractive point is marked with a
black point. The upper plot defines the geometry and the
notations.}
\end{figure}
	In this section we illustrate the results of the last sections
with several examples. We study mainly a billiard consisting of a
wedge of opening angle $\gamma$ whose two edges are connected by an
arc of constant radius of curvature $R$. The angles between arc and
wedge are chosen to be $\pi/3$ on both sides.  If $h$ denotes the
``height" of this billiard (see Fig.\ 6) then $R=h \sin(\gamma/2)
(\sin(\gamma/2)-1/2)^{-1}$. This billiard has only one diffractive
corner (at point $S$ of Fig.\ 6) and the curvature ensures that the
shortest diffractive and periodic orbits have $\mbox{Tr}\, M\neq\pm 2$
(they are displayed in Fig.\ 6). In the following we call this
billiard a ``rounded triangle ($\pi/3,\pi/3,\gamma$)".

\

	For numerical convenience we restrict ourselves to angles of
the form $\gamma=p\,\pi/q$ with $(p,q)\in\mbox{I$\!$N}^2$. The quantum
energies are determined by expanding the wave functions around point
$S$ in ``partial waves" which are Bessel functions times a sinusoidal
function of the angle:

\begin{equation}\label{ex1} \psi(r,\theta) = \sum_{n=1}^{n_{max}} J_{n
 q/p}(k r) \, \sin({\displaystyle n q\over\displaystyle p}\,\theta )
\; .\end{equation}

	Eq.\ (\ref{ex1}) automatically fulfills the Dirichlet
condition on the straight faces of the billiard. The boundary
condition on the arc opposite to $S$ is enforced in a manner identical
to the improved point matching method presented in \cite{Schm91}. This
results in a secular equation whose solutions are the eigen levels of
the system. We have tested the numerical stability of our procedure by
varying the number $n_{max}$ of partial waves included in the
expansion (typically $n_{max} \approx \, \mbox{nint} \, [p k h/q] \,
$). For each of the values of $\gamma$ studied below we have computed
the first 2000 eigenlevels and we have checked that they were
determined with an accuracy of the order of 1/1000 of the mean level
spacing.

\

	In order to visualize the importance of periodic and
diffractive orbits we study in the following the regularized Fourier
transform of the level density:

\begin{equation}\label{ex2} F(x)=\int_0^{k_{max}}\sqrt{k}\, 
\mbox{\Large e}^{\displaystyle i k x -\alpha k^2} d(k) \, dk \;
.\end{equation}

	$k_{max}$ is the last eigenvalue computed numerically and we
take here $\alpha=10/k_{max}^2$. The multiplicative factor $\sqrt{k}$
in (\ref{ex2}) is included in order to cancel the singularity at $k=0$
of a contribution of type (\ref{dr1}). $F(x)$ is denoted
$F_{\scriptscriptstyle QM} (x)$ if we use in (\ref{ex2}) the exact
quantum spectrum. It is denoted $F_{\scriptscriptstyle UA} (x)$ (resp.
$F_{\scriptscriptstyle GTD} (x)$) when Eq.\ (\ref{z34}) (resp.\
(\ref{dr1})) is used together with the periodic orbit contributions
(\ref{U23}).

\subsection{$\gamma$ near $\pi/2$}

	This case is relevant to the discussion 4.2 above. For
$\gamma<\pi/2$ the shortest orbit is a periodic one and has length
$L_1=2h\sin\gamma$. It disappears as soon as $\gamma>\pi/2 \,$. For
$\gamma\neq\pi/2$ one also has a diffractive orbit of length $L_2=2h$
(see Fig.\ 6). When $\gamma=\pi/2$ these orbits coalesce and give a
single periodic orbit of length $2h$. Their contribution to the level
density is continuous at $\gamma=\pi/2$ as explained above: the
contribution of $L_2$ is discontinuous (cf.\ Eq.\ (\ref{dr2})) and
this exactly cancels the discontinuity due to the disappearance of the
orbit $L_1$ and its time reverse.

\begin{figure}[thbp]
\begin{center}
\mbox{\epsfig{file=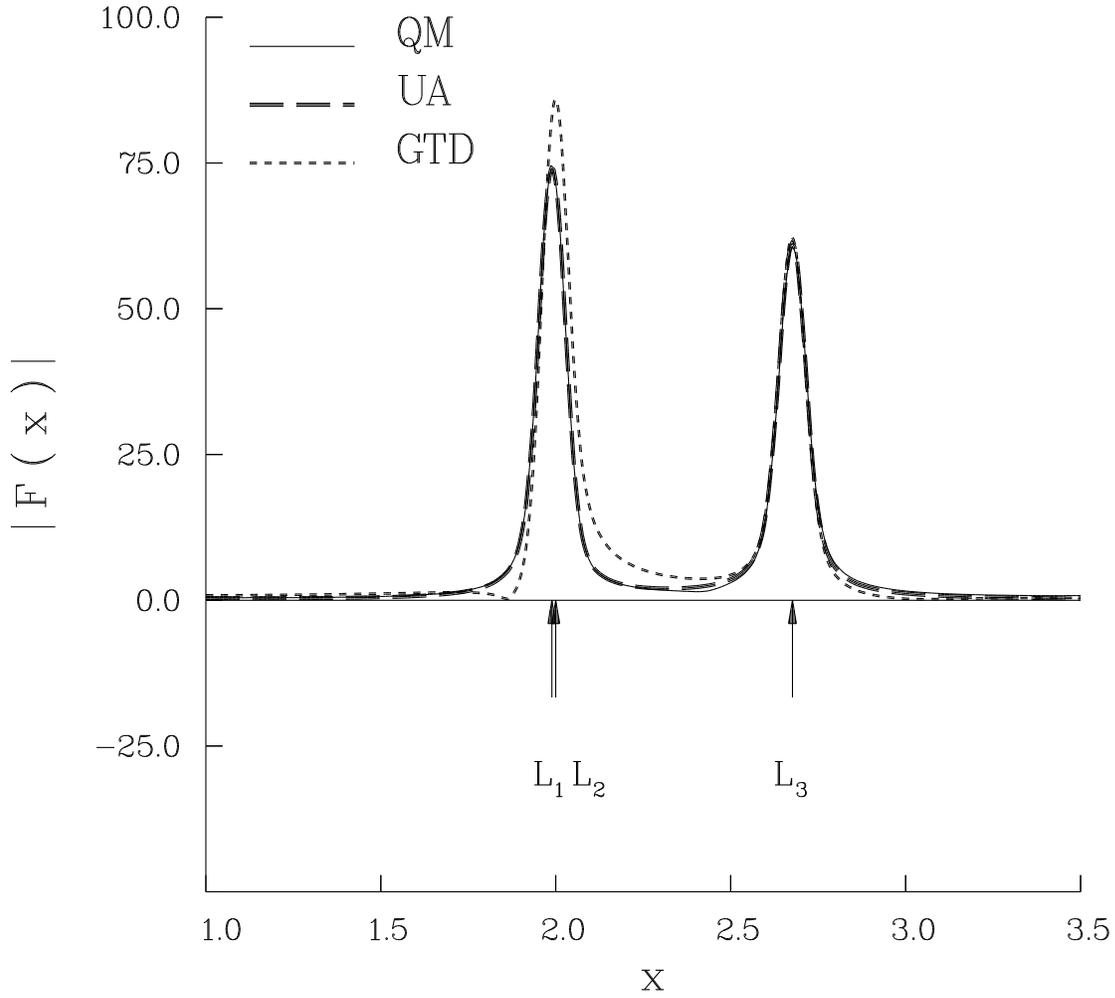,width=15cm,
bbllx=0pt, bblly=60pt, bburx=530pt, bbury=530pt,clip=}}
\end{center}
\caption{Modulus $|F(x)|$ of the Fourier transform of the
level density -- see Eq.\ (\protect\ref{ex2}) -- for the
rounded triangle ($\pi/3,\pi/3,\gamma=7\pi/15$). The solid
line corresponds to $F_{\protect\scriptscriptstyle QM}(x)$,
the long dashed line to $F_{\protect\scriptscriptstyle UA}(x)$
and the short dashed line to
$F_{\protect\scriptscriptstyle GTD}(x)$ (see the text).
The arrows mark the lengths of the diffractive and periodic orbits.
The scale of lengths and wave-vectors is fixed by taking $h=1$.}
\end{figure}
\begin{figure}[thbp]
\begin{center}
\mbox{\epsfig{file=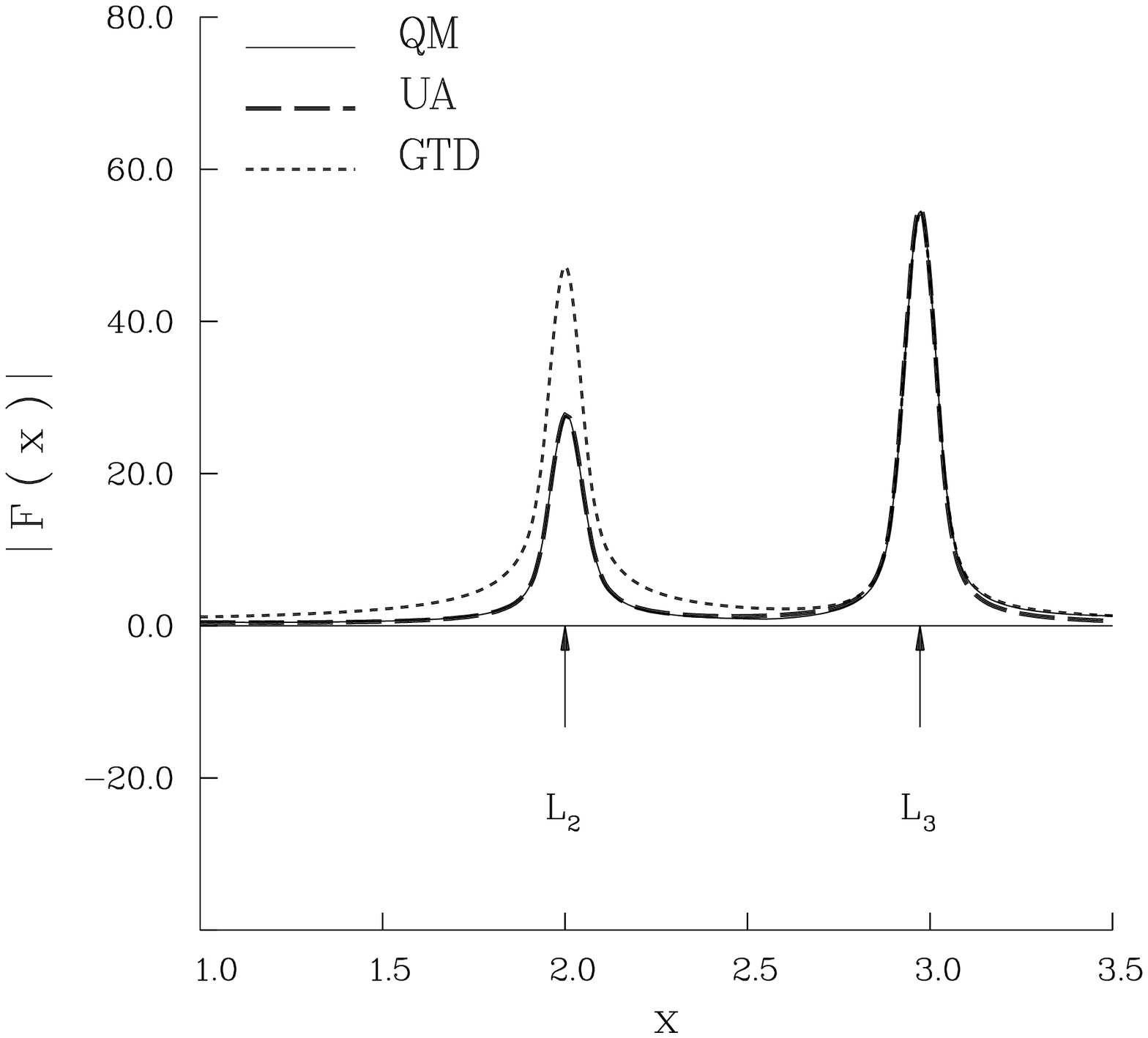,width=15cm,
bbllx=0pt, bblly=60pt, bburx=530pt, bbury=530pt,clip=}}
\end{center}
\caption{Same as Fig.\ 7 for $\gamma=8\pi/15$.}
\end{figure}
	We determined the spectrum numerically for $\gamma=7\pi/15$
and $8\pi/15$. The corresponding moduli of the Fourier transform
$|F(x)|$ are plotted in Fig.\ 7 and 8. In this figures the solid lines
correspond to $|F_{\scriptscriptstyle QM} (x)|$, the long dashed lines
to $|F_{\scriptscriptstyle UA} (x)|$ and the short dashed line to
$|F_{\scriptscriptstyle GTD} (x)|$. The lengths of the included orbits
are marked with back arrows. One notices the failure of GTD and the
excellent agreement of approximation (\ref{z34}) with the exact result
(the agreement remains equally good when plotting the real and
imaginary part of $F(x)$). As stated above, it can be seen that in the
vicinity of an optical boundary diffractive and periodic orbits
contribute in the same order to the level density.

\subsection{$\gamma$ near $\pi$}

	This case also pertains to the discussion of Sec.\ 4.2, but
now in the vicinity of $\gamma=\pi/(2p+1)$. Again the diffractive
orbit ensures continuity of semiclassical mechanics when $\gamma=\pi$:
the two periodic orbits $L_4$ and $L_4'$ of Fig.\ 6 disappear as soon
as $\gamma < \pi$, and the contribution of $L_2$ is discontinuous at
$\gamma=\pi$, but the joint contribution is continuous. Here we
computed numerically the levels for $\gamma=19\pi/20$ and present the
results for $|F(x)|$ and $\Re \{F(x)\}$ in Fig.\ 9. Again one can
verify the failure of the geometrical theory of diffraction and the
excellent agreement between (\ref{z34}) and the quantum result.
\begin{figure}[thbp]
\begin{center}
\mbox{\epsfig{file=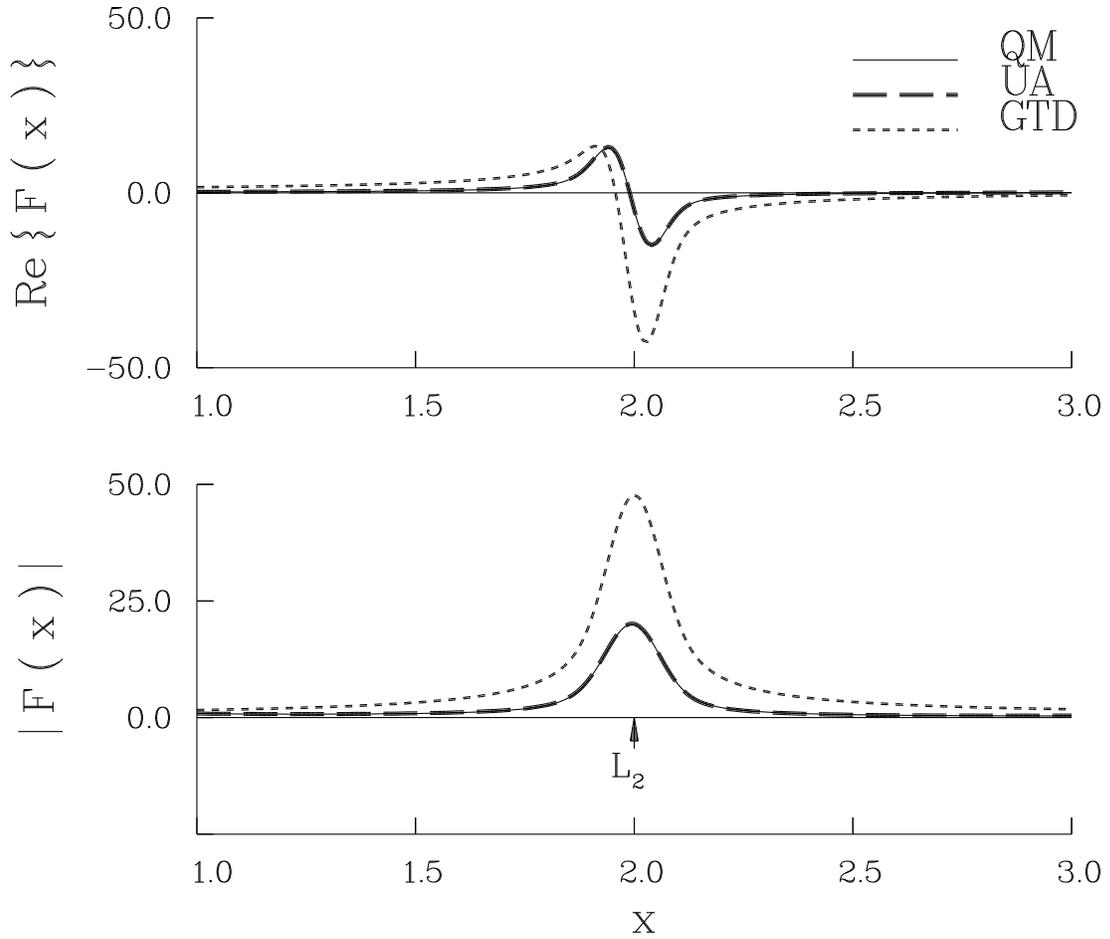,width=15cm,
bbllx=0pt, bblly=60pt, bburx=530pt, bbury=502pt,clip=}}
\end{center}
\caption{Same as Fig.\ 7 for $\gamma=19\pi/20$. We plot here
also $\protect\Re \{F(x)\}$ for illustrating the quality of
the agreement between the phases of 
$F_{\protect\scriptscriptstyle QM}$ and
$F_{\protect\scriptscriptstyle UA}$.}
\end{figure}

\subsection{The triangle ($\pi/4,\pi/6,7\pi/12$)}

	In this subsection we depart from the previous examples and
study the spectrum of a straight triangular billiard with angles
($\pi/4,\pi/6,7\pi/12$) which has one diffractive wedge
$\gamma=7\pi/12$. This billiard is of interest because (i) it allows
to compare the performances of the uniform approximation with GTD in a
regime where this last approximation is not inaccurate and also (ii)
because it provides an example where our approach is not completely
justified. Indeed in case of a polygonal billiard all the trajectories
have a monodromy matrix with $\mbox{Tr}\, M=\pm 2$ and this leads to a
divergence in (\ref{z34}). As noted in Sec.\ 4 this is linked to the
possible deformation of any diffractive orbit of the system considered
towards a family of periodic orbits. Fortunately, in the present case
the first diffractive orbits are far from any allowed family of
periodic orbits and we can evaluate the $K$-function relevant to the
divergent term in (\ref{z34}) with the asymptotic expansion
(\ref{f2}): this cancels the divergence. This was done on Fig.\ 10 for
the 3 first diffractive orbits of the system.
\begin{figure}[thbp]
\begin{center}
\mbox{\epsfig{file=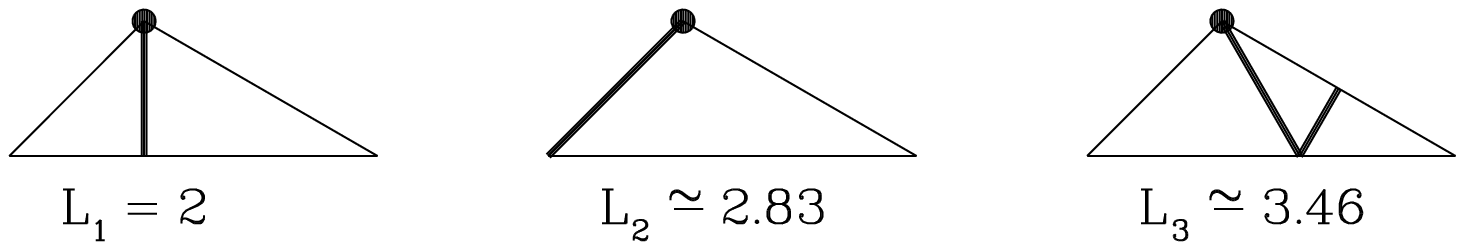,width=15cm,
bbllx=78pt, bblly=498pt, bburx=501pt, bbury=570pt,clip=}}
\end{center}
\vspace*{1.5cm}
\begin{center}
\mbox{\epsfig{file=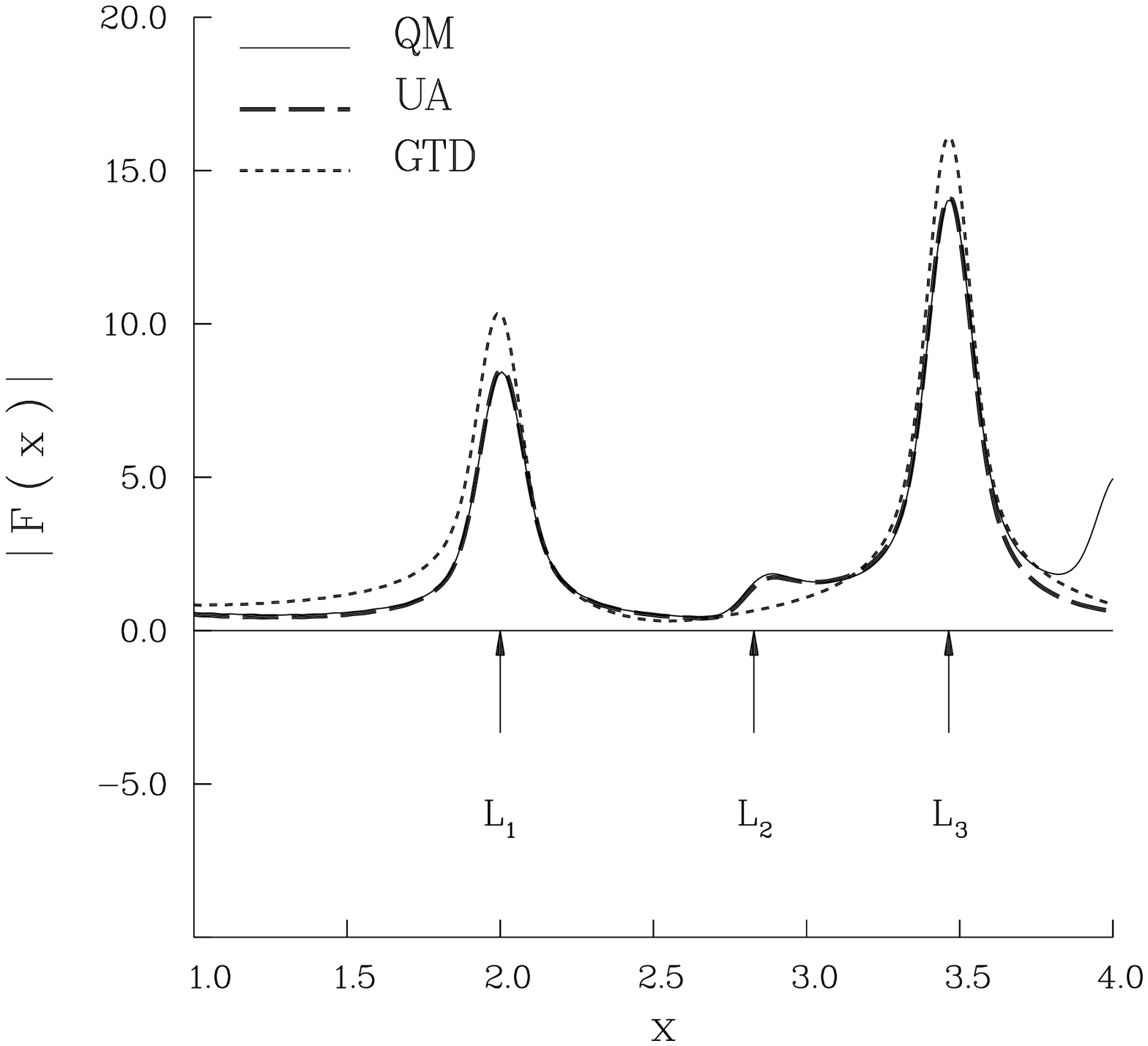,width=15cm,
bbllx=0pt, bblly=60pt, bburx=530pt, bbury=530pt,clip=}}
\end{center}
\caption{Same as Fig.\ 7 for the flat triangle
($\pi/4,\pi/6,7\pi/12$).  The upper part displays the shortest
orbits of the system (all three are diffractive, the classical
periodic orbits occur at greater lengths).}
\end{figure}
\begin{figure}[thbp]
\begin{center}
\mbox{\epsfig{file=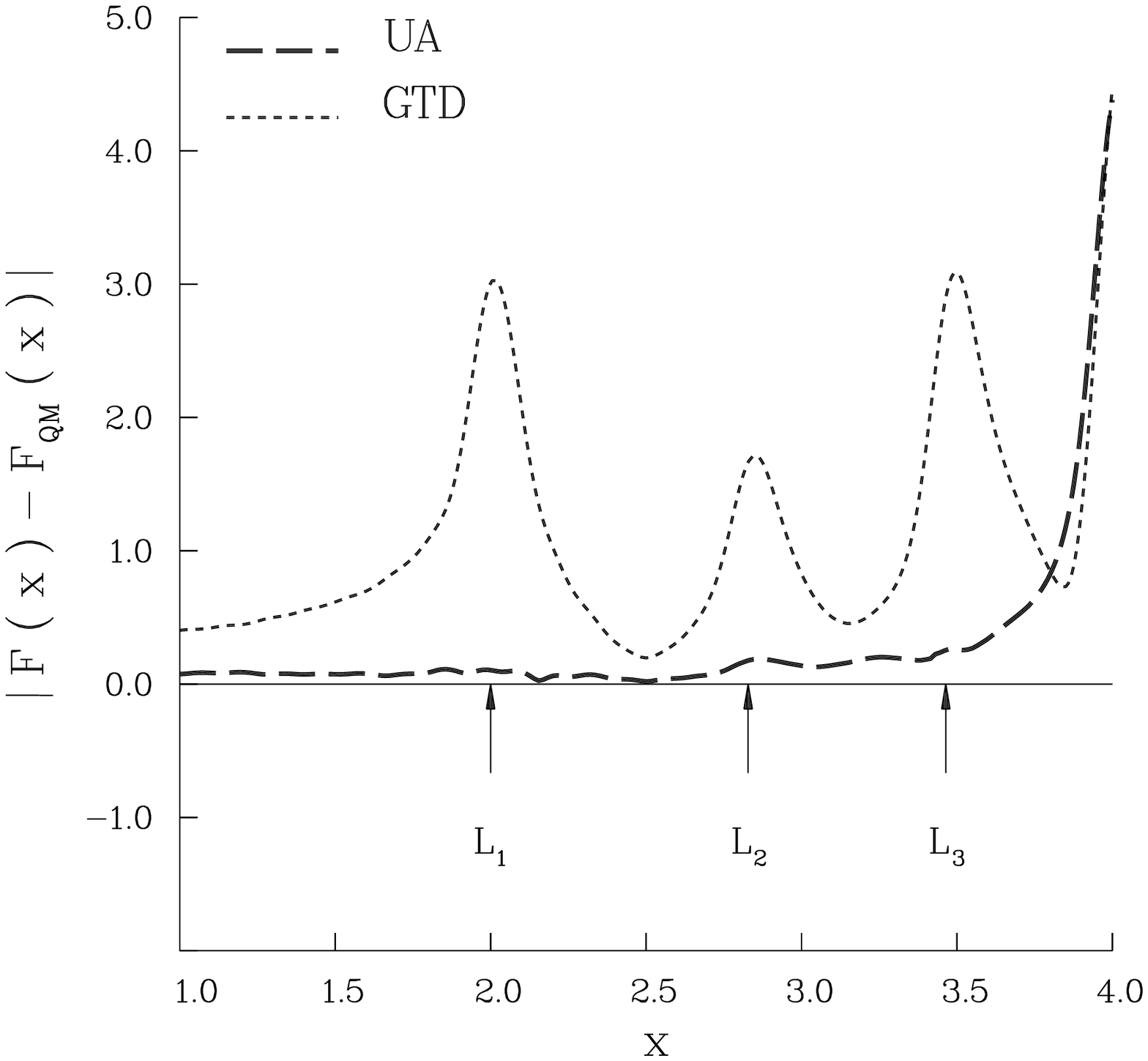,width=15cm,
bbllx=0pt, bblly=60pt, bburx=530pt, bbury=530pt,clip=}}
\end{center}
\caption{$|F_{\protect\scriptscriptstyle UA}(x)
 -F_{\protect\scriptscriptstyle QM}(x)|$ (long dashed line) and
$|F_{\protect\scriptscriptstyle GTD}(x)
 -F_{\protect\scriptscriptstyle QM}(x)|$ (short dashed line) for
the triangle ($\pi/4,\pi/6,7\pi/12$). We consider only the three
shortest orbits of the system. The following orbits are not taken
into account and this is the reason for the increasing errors in
vicinity of $x\protect\approx 4$.}
\end{figure}

	Again the agreement with the numerical result is excellent,
but here the geometrical theory of diffraction already gives a
sensible description. Note however that the small peak due to the
diffractive boundary orbit of length $L_2$ is ``missed" by GTD because
its diffractive coefficient (\ref{e8}) is zero (see the discussion in
2.2). The correct description of the peak was obtained by using half
the contribution (\ref{z34}) of a usual diffractive orbit.

	For a more detailed comparison we plot the moduli of the
differences $|F_{\scriptscriptstyle UA} (x) - F_{\scriptscriptstyle
QM} (x)|$ and $|F_{\scriptscriptstyle GTD} (x)- F_{\scriptscriptstyle
QM} (x)|$ in figure 11: even quite far from any optical boundary Eq.\
(\ref{z34}) supersedes the GTD result (\ref{dr1}). This plot
emphasizes the accuracy of Eq.\ (\ref{z34}) in cases slightly out of
its original range of application.

\section{Conclusion}

	In this paper we have studied the inclusion of diffractive
orbits in semiclassical trace formulae for billiards in which the
boundary has wedge-like singularities. In many cases the simple
geometrical theory of diffraction \cite{Keller} is inadequate,
especially if the energy is not very high. A consideration of the
mathematical structure of the exact Green function near a wedge
permits to remedy this shortcoming: it leads to a uniform
approximation of the Green function \cite{Kou74} which, in turn,
allows to derive contributions to the trace formula which properly
account for the role of isolated diffractive orbits in the quantum
spectrum (\ref{z34}). The formula was illustrated in several examples
and was shown to give excellent agreement with numerical data. Its
main feature is that it interpolates between the usual Gutzwiller
trace formula \cite{Gu90} and previous approaches relying on the
geometrical theory of diffraction \cite{Vat94,Pav95,Bru95}.

	Note also -- as a by-product of the present approach -- that
we derived a semiclassical composition law for Green functions for
billiard systems (\ref{U1},\ref{U14}) reminiscent to the semigroup
composition pro\-per\-ty of the propagator (see the discussion in
\cite{Gu90}). This allows to recover Gutzwiller's trace formula in a
simple fashion (cf. \ref{Compo}).  Similar laws can also be obtained
for the composition of diffractive and geometrical Green functions.

 	The present work suggests further developments: (i) the result
(\ref{z34}) might be extended to allow the treatment of diffractive
orbits in the vicinity of a family of periodic orbits; (ii) although
the inclusion of general multiple diffraction in a uniform formula
seems to be a difficult task, one may reasonably hope to include
double diffraction in the formalism (cf. \cite{Sch91}). (iii) Further
possible extensions concern the treatment of other types of
diffraction, like regions near curved wedges where surface diffraction
becomes important so that contributions from creeping and whispering
gallery orbits have to be included, or diffraction effects arising
from discontinuities of the curvature of the boundary like in the
stadium billiard.

	Finally we would like to emphasize the important role of
diffraction in semiclassical approaches. Diffractive and periodic
orbits are fundamentally different in the sense that the former are
not obtained via a systematic $\hbar$ expansion in the vicinity of
classical solutions of Hamilton's equations (they are rather linked to
discontinuities of the Hamiltonian flow).  However diffractive orbits
provide the first correction to the leading order in the trace
formula, with contributions typically of order $\sqrt{\hbar}$ smaller
than the contributions of isolated periodic orbits. Besides, in the
vicinity of optical boundaries the two types of orbit contribute with
approximately the same order to the trace formula. An image emerging
from our study (cf.  Sec.\ 4.2 and 4.3) is that diffractive orbits
allow to enforce semiclassically the continuity of wave mechanics in
the vicinity of discontinuities or bifurcations of classical
mechanics.

\bigskip \bigskip \noindent {\large \bf Acknowledgments}

\bigskip \noindent M.\,S.\ acknowledges financial support by the
Alexander von Humboldt-Stiftung and by the Deutsche
Forschungsgemeinschaft under contract No.\ DFG-Ste 241/7-1.  La
Division de Physique Th\'eorique de l'Institut de Physique Nucl\'eaire
est une unit\'e de recherche des universit\'es Paris XI et Paris VI
associ\'ee au CNRS.

\appendix

\section{The modified Fresnel function}
\setcounter{equation}{0}

	In this Appendix we define the modified Fresnel function
$K(z)$ used in the main text and list several of its properties.

	$\bullet$ The function $K(z)$ ($z\in\mbox{\large l$\!\!\!$C}$)
 is defined by:

\begin{equation}\label{f1} K(z)={\displaystyle \mbox{\Large 
e}^{\displaystyle -i(z^2+\pi/4)}\over\displaystyle\sqrt{\pi} }
\int_z^{\infty} \mbox{\Large e}^{\displaystyle i y^2} d y =
{\displaystyle \mbox{\Large e}^{\displaystyle -i
z^2}\over\displaystyle 2}\;\mbox{erfc}\;(\mbox{\Large
e}^{\displaystyle -i\pi/4}z) \; ,\end{equation}

\noindent where $\mbox{erfc}$ is the complementary error function (see
 e.\,g.\ \cite{Abra}). In (\ref{f1}) the path of integration is
 subject to the restriction $\mbox{arg}\; (y) \rightarrow \alpha$ with
 $0<\alpha<\pi/2$ as $y\rightarrow\infty$ along the path. $\alpha=0$
 and $\pi/2$ are permissible if $\mbox{Re}\; (i y^2)$ remains bounded
 to the right.

	The function $K$ has the following properties: $K(+\infty)=0$,
$K(0)=1/2$,

\begin{equation}\label{f1a} K(z)+K(-z) = \mbox{\Large 
e}^{\displaystyle -i z^2} \; , \end{equation}

\noindent and

\begin{equation}\label{f1b} \overline{K(\overline{z})}+K(-i z) = 
\mbox{\Large e}^{\displaystyle i z^2} \; , \end{equation}

\noindent where the bar denotes complex conjugation.

\

	$\bullet$ By successive integrations by parts one obtains the
following asymptotic expansion:

\begin{equation}\label{f2} K(z) = {\displaystyle \mbox{\Large 
e}^{\displaystyle i\pi/4}\over\displaystyle\ 2 z \sqrt{\pi} }
\sum_{n=0}^{+\infty} \left({\displaystyle 1\over\displaystyle
2}\right)_n \left({\displaystyle -i\over\displaystyle z^2}\right)^n
\quad \mbox{for}\quad |z|\rightarrow + \infty \quad\mbox{and}\quad
-\pi/4 < \; \mbox{arg}\;(z)\; < 3\pi/4 \; ,
\end{equation}

\noindent where $(1/2)_n=\Gamma(n+1/2)/\Gamma(1/2)=1\times 3\times
... \times(2n-1)/2^n$. In the region
$\mbox{arg}\;(z)\in\,]3\pi/4,7\pi/4[$ one obtains an asymptotic
expansion by combining Eqs.\ (\ref{f1a}) and (\ref{f2}).

\

	$\bullet$ The interest in the modified Fresnel function comes
from the following integral relation:

\begin{equation}\label{f3} \int_{-\infty}^{+\infty}\! d t \;
{\displaystyle \mbox{\Large e}^{\displaystyle -\beta
t^2}\over\displaystyle t-z} = 2\,i\tau\pi\, K\left( \tau\,\mbox{\Large
e}^{\displaystyle -i\pi/4}\sqrt{\displaystyle\beta } \,z \right) \;
,\end{equation}

\noindent where $\beta\in\mbox{I$\!$R}^+$, $z\in\mbox{\large
l$\!\!\!$C}$ and $\tau = \,\mbox{sign} \,\left(\,\Im\,(z) \right)$.

	Hence the function $K$ allows to generalize the steepest
descent method to cases where poles appear in the integrand. As
explained in the text (Sec.\ 2) this corresponds -- in the Sommerfeld
solution of the diffraction problem -- to the occurrence of
diffractive orbits near classical trajectories. We will not prove Eq.\
(\ref{f3}) here, it can be done easily by noting that
$(t-z)^{-1}=i\tau\int_0^{+\infty}\exp [i\tau(z-t)x]dx $ (cf. the
evaluation of integral (\ref{B7}) in \ref{Diffint}).

\section{Composition law for Green functions} \label{Compo}
\setcounter{equation}{0}

	In this Appendix we derive a simple semiclassical composition
law for Green functions which is expressed by integrals over the
boundary $\partial {\cal B}$ of the billiard.  Although the formulae
established below are simple and natural from the point of view of
Balian and Bloch's multiple reflexion expansion \cite{BB70}, to our
knowledge they have not been clearly stated in the literature. The
composition law can be used in order to simplify expressions obtained
from the boundary element method (cf. (\ref{z10})). We first prove the
semiclassical version and then give the exact formulation of this law
due to Balian and Bloch.  We further show that it allows to derive
Gutzwiller's trace formula in a straightforward manner.

We assume in this Appendix that the boundary $\partial {\cal B}$ is
smooth everywhere. The semiclassical version of the composition law
has the form

\begin{equation} \label{U1}
(-2)^n \int_{\partial {\cal B}} \! d s_1 \dots d s_n \,
                        G_0(\vec{r}_1,\vec{r}\,',E) \,
                        \partial_{\hat{n}_1} \,
                        G_0(\vec{r}_2,\vec{r}_1,E) \dots
                        \partial_{\hat{n}_n} \,
                        G_0(\vec{r},\vec{r}_n,E) \approx G^{(n)}_{sc}
                        (\vec{r},\vec{r}\,',E) \; .
\end{equation}

\noindent where $\vec{r}_i = \vec{r}(s_i)$.  The approximate sign
signifies that the evaluation is done by approximating the free Green
function $G_0$ by its leading asymptotic term for large argument and
evaluating the integrals in stationary phase approximation.  The
function $G^{(n)}_{sc} (\vec{r},\vec{r}\,',E)$ on the right-hand side
of Eq.\ (\ref{U1}) is the part of the semiclassical Green function
from all trajectories with $n$ bounces on the boundary between
$\vec{r}\,'$ and $\vec{r}$

\begin{equation} \label{U2}
G^{(n)}_{sc} (\vec{r},\vec{r}\,',E) = \sum_{\xi_n} \frac{1}{\sqrt{8
\pi k |M_{12}^{(n)}|}} \exp\{ i k l^{(n)} - i \frac{\pi}{2} \nu^{(n)}
- i \frac{3 \pi}{4} \} \; .
\end{equation}

	Here $l^{(n)}$ denotes the length of the trajectory,
$\nu^{(n)}$ is the number of conjugate points from $\vec{r}\,'$ to
$\vec{r}$ plus twice the number of reflections on the boundary, and
$M^{(n)}$ is the stability matrix for unit energy. An index $\xi_n$ of
the above quantities has been omitted in order to simplify the
notations.

Eq.\ (\ref{U1}) is proven by mathematical induction.  For $n=0$ it is
correct since

\begin{equation} \label{U3}
G_0(\vec{r},\vec{r}\,',E) \approx G^{(0)}_{sc} (\vec{r},\vec{r}\,',E)
\; ,
\end{equation}

\noindent and one has to show that

\begin{equation} \label{U4}
I := (-2) \int_{\partial {\cal B}} \! d s_1 \,
G_{sc}^{(n)}(\vec{r}_1,\vec{r}\,',E) \, \partial_{\hat{n}_1} \,
G_0(\vec{r},\vec{r}_1,E) \approx G^{(n+1)}_{sc} (\vec{r},\vec{r}\,',E)
\; .
\end{equation}

We will use the following notation at a point $\vec{r}_i$ of the
boundary: primed quantities correspond to the incoming trajectory and
unprimed quantities to the outgoing trajectory. We will denote the
momentum of a classical trajectory by a vector $\vec{p}$ of modulus
$k$ whose direction is the direction of propagation of the classical
particle. The momentum of an outgoing trajectory is $\vec{p}_i$ and
$\alpha_i$ is the angle between the normal vector $\hat{n}_i$ of the
boundary (which points outside) and $-\vec{p}_i$. The momentum of an
incoming trajectory is $\vec{p}_i'$, and $\alpha_i'$ is the angle
between $\hat{n}_i$ and $\vec{p}_i'$. For this choice $\alpha_i'$ and
$\alpha_i$ both lie in the interval between $-\pi/2$ and $\pi/2$. In
terms of the local coordinate systems of the trajectories with
coordinates parallel and perpendicular to the trajectory, the
tangential and normal vectors of the boundary can be written as:

\begin{equation} \label{U5}
\begin{array}{rrr}
\hat{n}_i & = - \cos \alpha_i \, \hat{e}_\parallel + \sin \alpha_i \,
              \hat{e}_\perp & = \cos \alpha_i' \, \hat{e}_\parallel' -
              \sin \alpha_i' \, \hat{e}_\perp' \; , \\ \hat{t}_i & = -
              \sin \alpha_i \, \hat{e}_\parallel - \cos \alpha_i \,
              \hat{e}_\perp & = \sin \alpha_i' \, \hat{e}_\parallel' +
              \cos \alpha_i' \, \hat{e}_\perp' \; .
\end{array}
\end{equation}

We continue by evaluating the integral in Eq.\ (\ref{U4}) using the
stationary phase approximation. The normal derivative of the Green
function is given in leading semiclassical order by

\begin{equation} \label{U6}
\partial_{\hat{n}_i} \, G_0(\vec{r}_{i+1},\vec{r}_i,E) \approx - i\,
\hat{n}_i \cdot \vec{p}_i \, G_{sc}^{(0)}(\vec{r}_{i+1},\vec{r}_i,E) =
i \, k \, \cos \alpha_i \, G_{sc}^{(0)}(\vec{r}_{i+1},\vec{r}_i,E) \;
,
\end{equation}

\noindent and in (\ref{U4}) the stationary points are determined by
the condition

\begin{equation} \label{U7}
0 = \frac{d}{d s_1} [l^{(0)} (\vec{r},\vec{r}_1) + l^{(n)}
(\vec{r}_1,\vec{r}\,')] = \hat{t} \cdot [ - \frac{\vec{p}_1}{k} +
\frac{\vec{p}_1\,'}{k}] = \sin \alpha_1 + \sin \alpha_1' \; ,
\end{equation}

\noindent i.\,e.\ by $\alpha_1 = -\alpha_1'$, which is the condition
for elastic reflection. The sum over all stationary points thus
expresses the integral $I$ by a sum over all trajectories with $n+1$
reflections on the boundary. In Eq.\ (\ref{U7}) and in the following
the length is given two arguments when it is necessary to specify the
starting and end point of the trajectory.

For the determination of the second derivatives of the lengths at a
boundary point $s_i$ one has to evaluate the derivatives of the angles
$\alpha_i$ and $\alpha_i'$ which consist of two parts. One from the
change of the normal vector with $s_i$ and one from the change of the
direction of the trajectories.

\begin{equation} \label{U8}
\begin{array}{llll}
\frac{\displaystyle d^2 l^{(n)}
(\vec{r}_i,\vec{r}_{i-1})}{\displaystyle d s_i^2} & = \cos \alpha_i'
\frac{\displaystyle d \alpha_i'}{\displaystyle d s_i} & =
\cos\alpha_i' \left( -\frac{\displaystyle 1}{\displaystyle R_i} +
\frac{\displaystyle\cos \alpha_i'}{\displaystyle k}
\frac{\displaystyle d p_{i_\perp}'}{\displaystyle d q_{i_\perp}'}
\right) & = -\frac{\displaystyle \cos \alpha_i'}{\displaystyle R_i} +
\frac{\displaystyle \cos^2 \alpha_i' \, M_{22}^{(n)}}{\displaystyle
M_{12}^{(n)}} \; ,\\ & & & \\ \frac{\displaystyle d^2 l^{(n)}
(\vec{r}_{i+1},\vec{r}_{i})}{\displaystyle d s_i^2} & = \cos \alpha_i
\frac{\displaystyle d \alpha_i}{\displaystyle d s_i} & = \cos\alpha_i
\left( -\frac{\displaystyle 1}{\displaystyle R_i} -
\frac{\displaystyle\cos \alpha_i}{\displaystyle k} \frac{\displaystyle
d p_{i_\perp}}{\displaystyle d q_{i_\perp}} \right) & =
-\frac{\displaystyle \cos \alpha_i}{\displaystyle R_i} +
\frac{\displaystyle \cos^2 \alpha_i \, M_{11}^{(n)}}{\displaystyle
M_{12}^{(n)}} \; .
\end{array}
\end{equation}

	At a stationary point it follows with these relations that

\begin{equation} \label{U9}
\frac{d^2}{d s_1^2} [l^{(0)} (\vec{r},\vec{r}_1) + l^{(n)}
(\vec{r}_1,\vec{r}\,')] = -\frac{\cos^2 \alpha_1 \, M_{12}^{(n+1)}
}{M_{12}^{(0)} M_{12}^{(n)}} \; ,
\end{equation}

\noindent where $M^{(n+1)} = M^{(0)} \, B_1 \, M^{(n)}$ and the
matrices $M^{(0)}$ and $B_1$ are given by

\begin{equation} \label{U10}
M^{(0)} = \left( \begin{array}{cc} 1 & l^{(0)} \\ 0 & 1 \end{array}
\right) \; , \; \; \; \; B_1 = \left( \begin{array}{cc} -1 & 0 \\
\frac{2}{R_1 \cos \alpha_1} & -1
\end{array} \right) \; .
\end{equation}

        The matrices $M$ and $B$ correspond to the linearized flow
near the considered trajectory. Note that our definition is slightly
different from usual conventions (see e.\,g.\ \cite{Gu90,Cre90}):
considering that here $|\vec{p} |=k$, the $M_{12}$ (resp.\ $M_{21}$)
matrix element would be generally divided (resp.\ multiplied) by
$k$. Here we work with the stability matrix at unit energy: this
choice is connected to the scaling property of the dynamics in
billiard systems. It does not affect the trace and the determinant of
the matrix and allows to have energy independent matrix elements with
a simple geometrical meaning.

	Now the stationary phase approximation for the integral in
Eq.\ (\ref{U4}) is carried out and results in:

\begin{eqnarray} \label{U12}
I &\approx& \sum_{\xi_{n+1}} \frac{\cos \alpha_1 \, \exp \{ i k
l^{(n+1)} - i \frac{\pi}{2} \nu^{(n)} \} }{4 \pi \sqrt{|M_{12}^{(0)}
M_{12}^{(n)}|}} \int_{\partial {\cal B}} \! d s_1 \, \exp \{- i k
\frac{\cos^2 \alpha_1 \, M_{12}^{(n+1)} }{ 2 M_{12}^{(0)}
M_{12}^{(n)}} s^2_1 \} \nonumber \\ &\approx& \sum_{\xi_{n+1}}
\frac{1}{ \sqrt{8 \pi k |M_{12}^{(n+1)}|}} \exp \{ i k l^{(n+1)} - i
\frac{\pi}{2} \nu^{(n+1)} - i \frac{3 \pi}{4} \} \; ,
\end{eqnarray}

\noindent where $l^{(n+1)} = l^{(0)} + l^{(n)}$ and

\begin{equation} \label{U13}
\nu^{(n+1)} = \nu^{(n)} + 2 + \left\{ \begin{array}{rcl} 1 & \mbox{if}
& \mbox{sign}(M_{12}^{(n+1)}) = \mbox{sign}(M_{12}^{(n)}) \; ,\\ 0 &
\mbox{if} & \mbox{sign}(M_{12}^{(n+1)}) \neq
\mbox{sign}(M_{12}^{(n)})\; .
\end{array} \right.
\end{equation}

	Eq.\ (\ref{U13}) coincides with the expected definition of the
 Maslov index: $\nu^{(n+1)}$ is the number of conjugate points from
 $\vec{r}\,'$ to $\vec{r}$ plus twice the number of reflections on the
 boundary; an additional conjugate point has occurred between
 $\vec{r}_1$ and $\vec{r}$ iff $\mbox{sign}(M_{12}^{(n+1)}) =
 \mbox{sign}(M_{12}^{(n)})$ (remember that there is one sign change
 due to the reflection on the boundary). This completes the proof of
 Eq.\ (\ref{U4}) and thus also of Eq.\ (\ref{U1}).

A further relation follows from the fact that the evaluation of the
integral in Eq.\ (\ref{U1}) does not depend on the order in which the
stationary phase approximations are carried out. Thus one can conclude
directly that

\begin{equation} \label{U14}
(-2) \int_{\partial {\cal B}} \! d s_1 \,
G_{sc}^{(n)}(\vec{r}_1,\vec{r}\,',E) \, \partial_{\hat{n}_1} \,
G_{sc}^{(m)}(\vec{r},\vec{r}_1,E) \approx G^{(n+m+1)}_{sc}
(\vec{r},\vec{r}\,',E) \; .
\end{equation}

Equations (\ref{U1}) and (\ref{U14}) were derived in the semiclassical
approximation by evaluating the boun\-da\-ry integrals only locally in
the vicinity of stationary points. For that reason the same
composition law can be applied in order to obtain the contributions of
the geometrical orbits in billiards with corners; this is done in Eq.\
(\ref{z9}).

We note that equations (\ref{U1}) and (\ref{U14}) are the
semiclassical versions of {\it exact} relations for the Green function
$G$ of a billiard system. These exact relations are obtained by a
multiple reflection expansion of the Green function $G$ \cite{BB70}:

\begin{equation} \label{U15}
G(\vec{r},\vec{r}\,',E) = \sum_{n=0}^\infty
G^{(n)}(\vec{r},\vec{r}\,',E)\; ,
\end{equation}

\noindent where

\begin{equation} \label{U16}
G^{(n)}(\vec{r},\vec{r}\,',E) = (-2)^n \int_{\partial {\cal B}} \! d
s_1 \dots d s_n \, G_0(\vec{r}_1,\vec{r}\,',E) \, \partial_{\hat{n}_1}
\, G_0(\vec{r}_2,\vec{r}_1,E) \dots \partial_{\hat{n}_n} \,
G_0(\vec{r},\vec{r}_n,E) \; ,
\end{equation}

\noindent and the equation analogous to (\ref{U14}) follows directly.

Finally, we show that Gutzwiller's trace formula can be obtained in a
straightforward way by using (\ref{U1}). From the boundary element
method one obtains

\begin{equation} \label{U17}
d(k) = \bar{d}(k) + \frac{1}{\pi} \Im \sum_{n=1}^\infty \frac{1}{n}
\frac{d}{d k} \mbox{Tr}\, \hat{Q}^n (k)\; ,
\end{equation}

\noindent where

\begin{equation} \label{U18}
\mbox{Tr}\, \hat{Q}^n (k) = (-2)^n \int_{\partial {\cal B}} \! d s_1
\dots d s_n \, \partial_{\hat{n}_1} \, G_0(\vec{r}_2,\vec{r}_1,E)
\partial_{\hat{n}_2} \, G_0(\vec{r}_3,\vec{r}_2,E) \dots
\partial_{\hat{n}_n} \, G_0(\vec{r}_1,\vec{r}_n,E) \; .
\end{equation}

	With Eq.\ (\ref{U1}) it follows that

\begin{eqnarray} \label{U19}
d(k) &\approx& \bar{d}(k) - \frac{2}{\pi} \, \frac{d}{d k} \Im
\sum_{n=1}^\infty \frac{1}{n} \int_{\partial {\cal B}} \! d s \,
\partial_{\hat{n}'} \, G_{sc}^{(n-1)}(\vec{r},\vec{r}\,',E)
|_{\vec{r}=\vec{r}\,'} \nonumber \\ &\approx& \bar{d}(k) -
\frac{2}{\pi} \Re \sum_{n=2}^\infty \frac{1}{n} \frac{k \, l^{(n-1)}
\cos \alpha}{\sqrt{8 \pi k |M_{12}^{(n)}|}} \int_{\partial {\cal B}}
\! d s \exp \{ i k l^{(n-1)}(\vec{r},\vec{r}) - i \frac{\pi}{2}
\nu^{(n-1)} - i \frac{\pi}{4} \} \; .
\end{eqnarray}

	The stationary phase condition is again given by $\sin \alpha
= - \sin \alpha'$ and thus the integral yields contributions from
periodic orbits with $n$ specular reflections on the boundary. More
accurately, it gives $n/r_{po}$ (identical) contributions for every
periodic orbit where $r_{po}$ is the repetition number of the orbit,
since there are $n/r_{po}$ different starting positions
$\vec{r}=\vec{r}\,'$ on $\partial {\cal B}$.

	The derivatives of the angles $\alpha$ and $\alpha'$ now have
additional contributions since both initial and final points of the
trajectory are changed by varying $s$

\begin{equation} \label{U20}
\begin{array}{lll}
\frac{\displaystyle d \alpha'}{\displaystyle d s} & = -
\frac{\displaystyle 1}{\displaystyle R} + \left. \frac{\displaystyle
\cos \alpha'}{\displaystyle k} \frac{\displaystyle d
p_\perp'}{\displaystyle d q_\perp'} \right|_{q_\perp} -
\left. \frac{\displaystyle \cos \alpha}{\displaystyle k}
\frac{\displaystyle d p_\perp'}{\displaystyle d q_\perp }
\right|_{q_\perp'} & = - \frac{\displaystyle 1}{\displaystyle R} +
\frac{\displaystyle \cos \alpha' \, M_{22}^{(n-1)} + \cos \alpha
}{\displaystyle M_{12}^{(n-1)}} \; , \\ & & \\ \frac{\displaystyle d
\alpha}{\displaystyle d s} & = -\frac{\displaystyle 1}{\displaystyle
R} - \left. \frac{\displaystyle \cos \alpha}{\displaystyle k}
\frac{\displaystyle d p_\perp}{\displaystyle d q_\perp }
\right|_{q_\perp'} + \left. \frac{\displaystyle \cos
\alpha'}{\displaystyle k} \frac{\displaystyle d p_\perp}{\displaystyle
d q_\perp'} \right|_{q_\perp} & = - \frac{\displaystyle
1}{\displaystyle R} + \frac{\displaystyle \cos \alpha \,
M_{11}^{(n-1)} + \cos \alpha' }{\displaystyle M_{12}^{(n-1)}} \; .
\end{array}
\end{equation}

It then follows at a stationary point that

\begin{equation} \label{U21}
\frac{d^2}{d s^2} l^{(n-1)} (\vec{r},\vec{r}) = - \frac{\cos^2 \alpha
\, (\mbox{Tr}\, M^{(n)}_{po} - 2) }{(M_{po}^{(n)})_{12}} \; ,
\end{equation}

\noindent where $M_{po}^{(n)} = B_1 M^{(n-1)}$.

and the stationary phase approximation results in

\begin{equation} \label{U23}
d(k) = \bar{d}(k) + \frac{1}{\pi} \sum_{n=1}^\infty \sum_{\xi_{n,po}}
\frac{l^{(n)}_{po}}{r_{po} \, \sqrt{|\mbox{Tr}\, M^{(n)}_{po} - 2|}}
\cos \{ k l^{(n)}_{po} - \frac{\pi}{2} \mu^{(n)}_{po} \} \; ,
\end{equation}

\noindent where

\begin{equation} \label{U24}
\mu^{(n)}_{po} = \nu^{(n-1)} + 2 + \left\{
\begin{array}{rcl}
0 & \mbox{if} & (M_{po}^{(n)})_{12} / (\mbox{Tr}\, M^{(n)}_{po} - 2) >
 0 \; ,\\ 1 & \mbox{if} & (M_{po}^{(n)})_{12} / (\mbox{Tr}\,
 M^{(n)}_{po} - 2) < 0 \; .
\end{array} \right.
\end{equation}

	Note that the derivation presented here has the same starting
point as Ref.\ \cite{Har92}. But the composition law (\ref{U1})
permits to bypass the computation of large determinants of
\cite{Har92}.  Furthermore, it allows to keep track of the Maslov
indices (which were not derived in \cite{Har92}) in a simple way.

	Finally we add a remark on ghost contributions. In general,
the semiclassical approximation for the Green functions
$G^{(n)}(\vec{r},\vec{r}\,',E)$ can also contain contributions from
ghost trajectories that satisfy the stationary phase conditions but
have parts that are outside the billiard region.  These ghost orbits,
however, do not give a contribution to the level density $d(k)$ since
they cancel with ghost contributions from different $n$ or from
$\bar{d}(k)$ \cite{BB72,Har92,Alo94,Hesse}.

\section{Evaluation of a diffraction integral} \label{Diffint}
\setcounter{equation}{0}

In this appendix the integral

\begin{equation} \label{B1}
I = \int_{-\infty}^\infty \! ds \, \int_{-i\infty}^{i\infty} \! dz \,
    \frac{\mbox{\Large e}^{\displaystyle i a s^2 - i c z^2}}{z + s -
    s_0} \; ,
\end{equation}

\noindent is evaluated for positive $c$ and real non-vanishing $a$ and
 $s_0$.  This is the basic integral which appears in the derivation of
 the uniform approximation for diffractive contributions to the trace
 formula.

First the $z$-integral is rotated onto the real axis.  The rotation is
performed counter-clockwise: since $c>0$ this yields no contribution
from infinity. There are, however, poles of the integrand on the real
$z$ line.  We take them into account by giving to $s_0$ a small
imaginary part $s_0 \rightarrow s_0 + i \sigma_0 \varepsilon$ and
consider the limit $\varepsilon \rightarrow 0$ in the end.  Here
$\sigma_0 = \mbox{sign}(s_0)$ and $\varepsilon > 0$.  For this choice
one obtains a pole contribution from the rotation of the $z$-integral
for those values of $s$ for which $(s_0 - s)$ has a different sign
than $s_0$. One obtains

\begin{equation} \label{B2}
\int_{-i\infty}^{i\infty} \! dz \, \frac{\mbox{\Large
e}^{\displaystyle - i c z^2}}{z + s - s_0} = \lim_{\varepsilon
\rightarrow 0} \int_{\infty}^{-\infty} \! dz \, \frac{\mbox{\Large
e}^{\displaystyle - i c z^2}}{z + s - s_0 - i \sigma_0 \varepsilon} +
2 \pi i \sigma_0 \, \mbox{\Large e}^{\displaystyle - i c (s-s_0)^2}
\Theta (\sigma_0(s - s_0)) \; .
\end{equation}

We consider now the two contributions of the r.h.s.\ of Eq.\
(\ref{B2}) to the integral in (\ref{B1}) separately, $I = I_0 + I_1$,
where $I_0$ contains the pole contribution and $I_1$ the contribution
from the rotated $z$-integral.  For $I_0$ we have

\begin{eqnarray} \label{B4}
I_0 &=& 2 \pi i \sigma_0 \int_{-\infty}^\infty \!  ds \, \Theta
        (\sigma_0(s - s_0)) \, \mbox{\Large e}^{\displaystyle i a s^2
        - i c (s - s_0)^2} \nonumber \\ &=& 2 \pi i \sigma_0
        \int_{|s_0|}^\infty \! ds \, \mbox{\Large e}^{\displaystyle i
        a s^2 - i c (s - |s_0|)^2} \nonumber \\ &=& \frac{i \pi
        \sqrt{\pi} \sigma_0}{\sqrt{-i(a-c)}} \, \exp\{ - \frac{i a c
        s_0^2}{a-c} \} \, \mbox{erfc} \{ \frac{-i a
        |s_0|}{\sqrt{-i(a-c)}} \} \; ,
\end{eqnarray}

\noindent where $\mbox{erfc}$ is the complementary error function (see
 e.\,g.\ \cite{Abra}). $I_1$ has the form

\begin{equation} \label{B5}
I_1 = - \lim_{\varepsilon \rightarrow 0} \int_{-\infty}^\infty \! ds
      \, \int_{-\infty}^\infty \! dz \, \frac{\mbox{\Large
      e}^{\displaystyle i a s^2 - i c z^2}}{ z + s - s_0 - i \sigma_0
      \varepsilon} \; .
\end{equation}

	By a linear transformation of the variables

\begin{equation} \label{B6}
u = z + s \; \; \; , \; v = \frac{c z}{a - c} + \frac{a s}{a-c} \; ,
\end{equation}

\noindent the double integral splits into a product of two single
integrals

\begin{equation} \label{B7}
I_1 = - \lim_{\varepsilon \rightarrow 0} \int_{-\infty}^\infty \! d v
        \, \mbox{\Large e}^{\displaystyle i (a - c) v^2}
        \int_{-\infty}^\infty \! d u \, \frac{\mbox{\Large
        e}^{\displaystyle - i \frac{a c}{a-c} u^2} }{u - s_0 - i
        \sigma_0 \varepsilon} \; .
\end{equation}

	In (\ref{B7}) the integral over $v$ can be computed easily.
Furthermore, the denominator in the $u$-integral can be expressed in
terms of an integral

\begin{eqnarray} \label{B8}
I_1 &=& - i \sigma_0 \sqrt{\frac{\pi}{-i(a-c)}} \lim_{\varepsilon
\rightarrow 0} \int_{-\infty}^\infty \! d u \, \int_0^\infty \! d w \;
\mbox{\Large e}^{\displaystyle - i \frac{a c}{a-c} u^2 - i \sigma_0 w
(u - s_0 - i \sigma_0 \varepsilon)} \nonumber \\ &=& - i \sigma_0
\sqrt{\frac{\pi}{-i(a-c)}} \sqrt{\frac{\pi (a-c)}{i a c}}
\int_0^\infty \! d w \; \mbox{\Large e}^{\displaystyle i \frac{a-c}{4a
c} w^2 + i |s_0| w} \nonumber \\ &=& - \frac{i \pi \sqrt{\pi}
\sigma_0}{\sqrt{-i(a-c)}} \, \exp\{ - \frac{i a c s_0^2}{a-c} \} \,
\mbox{erfc} \{ - i |s_0| \sqrt{\frac{i a c}{a-c}} \} \; .
\end{eqnarray}

The whole result $I = I_0 + I_1$ is given by

\begin{equation} \label{B9}
I = \frac{i \pi \sqrt{\pi} \sigma_0}{\sqrt{-i(a-c)}} \, \exp\{ -
    \frac{i a c s_0^2}{a-c} \} \, \left[ \mbox{erfc} \{ \frac{-i a
    |s_0|}{\sqrt{-i(a-c)}} \} - \mbox{erfc} \{ - i |s_0| \sqrt{\frac{i
    a c}{a-c}} \} \right] \; .
\end{equation}

	It is convenient to rewrite this result in a form in which the
phases of the complex arguments of the error functions are always
between $-\pi/2$ and $\pi/2$. This can be done by considering all the
possible cases for the signs of $a$ and $(a-c)$ separately and using
the relation $\mbox{erfc} (z) = 2 - \mbox{erfc} (-z)$. The results for
the different cases can be combined again and written in the form

\begin{equation} \label{B10}
I = \frac{\tau \sigma_0 \pi \sqrt{\pi}}{\sqrt{|a-c|}} \, \mbox{\Large
    e}^{\displaystyle i \frac{\pi}{4} (1 + \sigma_a + \tau)} \exp\{ -
    \frac{i a c s_0^2}{a-c} \} \, \left[ \mbox{erfc} \{ \frac{|a
    s_0|}{\sqrt{i(a-c)}} \} - \mbox{erfc} \{ |s_0| \sqrt{\frac{a
    c}{i(a-c)}} \} \right] \; ,
\end{equation}

where $\sigma_a = \mbox{sign}(a)$ and $\tau = \mbox{sign}(a/(a-c))$.

\section{Curved wedges}
\setcounter{equation}{0}

In this section we discuss the effect of curved wedges on the
contributions of diffractive orbits to the level density.  The uniform
approximation (\ref{z34}) has been derived for a boundary with zero
curvature on both sides of the corner. It has to be modified for
curved wedges, otherwise the sum of diffractive and periodic orbit
contributions is not continuous any more as an optical boundary is
crossed. Additional complications can arise due to surface diffraction
effects, i.\,e.\ creeping orbit or whispering orbit contributions can
interfere with the diffractive orbit contributions. We will discuss
when these effects have to be taken into account, but we will modify
the uniform approximation only in those regions in which these
additional effects can be neglected.

The modified formula is derived by using a method of Ref.\
\cite{Kou74} for obtaining a uniform approximation for the Green
function in the case of a curved wedge (see also Ref.\
\cite{James}). We refer to the original references for a discussion of
this method and state here only the result which consists of a change
of the argument of the Fresnel function in (\ref{e17}) such that the
approximation is continuous across an optical boundary. For the
diffractive orbit contribution to the level density, this has the
consequence that only the stability matrix $M$ is changed in Eqs.\
(\ref{z32},\ref{z33},\ref{z34}): there are additional contributions
to $M$ from reflections on the curved boundary.

In order to discuss these modifications, we first list several
properties of geometrical orbits cor\-res\-pon\-ding to an optical
boundary which is specified by the values of $\sigma$, $\eta$ and
$n_{\sigma,\eta}$. In particular, we consider the trajectories which
contribute to the Green function and list for them the number of
reflections on the boundary and the side of the corner on which the
first reflection occurs. Furthermore, we give restrictions for the
numbers $n_{\sigma,\eta}$ which are implied by their definition.

\begin{itemize}
\item $\sigma = +1$, $\eta = +1$: $n_{\sigma,\eta} \geq 0$, $(2
n_{\sigma,\eta})$ reflections, first on the line $\theta = \gamma$.
\item $\sigma = +1$, $\eta = -1$: $n_{\sigma,\eta} \leq 0$, $(-2
n_{\sigma,\eta})$ reflections, first on the line $\theta = 0$.
\item $\sigma = -1$, $\eta = +1$: $n_{\sigma,\eta} \geq 0$, $(2
n_{\sigma,\eta}-1)$ reflections, first on the line $\theta =
\gamma$. If $n_{\sigma,\eta} = 0$ the optical boundary cannot be
reached. This case can occur only for $\gamma > \pi$.
\item $\sigma = -1$, $\eta = -1$: $n_{\sigma,\eta} \leq 1$, $(1-2
n_{\sigma,\eta})$ reflections, first on the line $\theta = 0$. The
optical boundary cannot be reached if $n_{\sigma,\eta} = 1$. This case
can occur only for $\gamma > \pi$.
\end{itemize}

	With these properties we can now discuss the modification of
the stability matrix $M$ in the case of curved wedges: $M$ then has an
additional contribution for every of the reflections mentioned
above. In the following we denote the limits of the radii of curvature
as the corner is approached from either side by $R_0$ and $R_\gamma$
where the first one corresponds to the side $\theta=0$ and the second
one to $\theta = \gamma$. Then $M$ has to be replaced by $BM$ where

\begin{equation} \label{DD1}
B = \left( \begin{array}{cc} 1 & 0 \\ -b & 1
\end{array} \right) \; .
\end{equation}
\noindent and
\begin{equation} \label{DD2}
b = \left\{ \begin{array}{lllll} \displaystyle
\sum_{j=1}^{n_{\sigma,\eta}} & \frac{\displaystyle 2}{\displaystyle
R_0 \sin(2j \gamma - \theta_1)} & +
\displaystyle\sum_{j=0}^{n_{\sigma,\eta}-1} & \frac{\displaystyle
2}{\displaystyle R_\gamma \sin((2j+1) \gamma - \theta_1)} & \mbox{if}
\; \sigma = +1, \eta = +1 \\ \displaystyle
\sum_{j=0}^{-n_{\sigma,\eta}-1} & \frac{\displaystyle 2}{\displaystyle
R_0 \sin( 2j \gamma + \theta_1)} & + \displaystyle
\sum_{j=1}^{-n_{\sigma,\eta}} & \frac{\displaystyle 2}{\displaystyle
R_\gamma \sin((2j-1) \gamma + \theta_1)} & \mbox{if} \; \sigma = +1,
\eta = -1 \\ \displaystyle\sum_{j=1}^{n_{\sigma,\eta}-1} &
\frac{\displaystyle 2}{\displaystyle R_0 \sin(2j \gamma - \theta_1)} &
+ \displaystyle\sum_{j=0}^{n_{\sigma,\eta}-1} & \frac{\displaystyle
2}{\displaystyle R_\gamma \sin((2j+1) \gamma - \theta_1)} & \mbox{if}
\; \sigma = -1, \eta = +1 \\
\displaystyle\sum_{j=0}^{-n_{\sigma,\eta}} & \frac{\displaystyle
2}{\displaystyle R_0 \sin( 2j \gamma + \theta_1)} & +
\displaystyle\sum_{j=1}^{-n_{\sigma,\eta}} & \frac{\displaystyle
2}{\displaystyle R_\gamma \sin((2j-1) \gamma + \theta_1)} & \mbox{if}
\; \sigma = -1, \eta = -1 \\
\end{array} \right.
\end{equation}

This approximation is only valid as long as all sine-functions in
(\ref{DD2}) are positive and not close to zero. The case of an almost
vanishing sine-function corresponds to almost grazing incidence on a
side of the corner.  Then surface diffraction effects become important
and interfere with the diffractive orbit contribution, and the uniform
approximation is no longer valid. In the case that some sine-functions
are negative and not small, the orbit is not close to an optical
boundary and the GTD-approximation can be used (it is the same as in
the case of non-curved wedges).

	There is a disadvantage of the definition of $B$ given above.
Since there are two possibilities for choosing $\theta_1$ and
$\theta_2$ corresponding to the two arms of a diffractive orbit in a
corner, it follows from (\ref{DD2}) that the uniform approximation is
not uniquely defined. (Note that both cases have to be checked for
deciding whether surface diffraction effects are important.)

	The non-uniqueness of the approximation is a direct
consequence of the fact that the uniform approximation for the Green
function of Ref.\ \cite{Kou74} is not symmetric under $\theta_1
\leftrightarrow \theta_2$. It is another example for the
non-uniqueness of uniform approximations (cf.  the discussion in Sec.\
2.3). However, as an optical boundary is approached both choices give
the same result as they should. Let us discuss in more detail the
difference between these two choices. It can be shown that
interchanging $\theta_1$ and $\theta_2$ amounts to evaluate
(\ref{DD2}) with $\theta_1 = \sigma \theta_2 + 2 n_{\sigma,\eta}
\gamma - \eta \pi$ instead of $\theta_1$.  This then directly suggests
a possible way by which this ambiguity can be removed, namely by
replacing $\theta_1$ in (\ref{DD2}) by the average of both values
which is $(\theta_1 + \sigma \theta_2 + 2 n_{\sigma,\eta} \gamma -
\eta \pi)/2$. As an optical boundary is approached this combination
again becomes identical to $\theta_1$.

\renewcommand{\baselinestretch}{.7}

\end{document}